%% file: OJA.tex
\documentclass[apj,twocolumn,numberedappendix]{openjournal}
\usepackage{amsmath}
\usepackage{booktabs}
\usepackage{multirow}
\usepackage{soul}
\usepackage{adjustbox}

\usepackage[dvipsnames]{xcolor} %added later for color

\newcommand{\Msun}{\mathrm{M}_\odot}	%Msun
\newcommand{\Rsun}{\mathrm{R}_\odot}	%Rsun
\newcommand{\Lsun}{\mathrm{L}_\odot}	%Lsun
\newcommand{\Teff}{T_\mathrm{eff}}	%Teff
\newcommand{\Mi}{M_\mathrm{i}}
\newcommand{\Xc}{X_\mathrm{c}}
\newcommand{\Mc}{M_\mathrm{c}}
\newcommand{\McTAMS}{M_\mathrm{c,TAMS}}

\newcommand{\Yc}{Y_\mathrm{c}}

\newcommand{\etac}{\eta_\mathrm{c}}
\newcommand{\etacHeF}{\eta_\mathrm{c,HeF}}

\renewcommand{\arraystretch}{1.2}  
\usepackage{listings}
\usepackage{hyperref}  % Load hyperref first
\usepackage{orcidlink} % Load orcidlink after hyperref
\usepackage{nameref}   % Load nameref last

\begin{document}

%\preprint{}

\title[A Binary Stellar Evolutionary Grid for the Giant Branch]{A Stellar Evolutionary Grid for Binary Population Synthesis: From the Main Sequence to Helium Ignition}

\author{Natalie R. Rees\,\orcidlink{0009-0004-7488-6542}$^{1}$}
\author{Robert G. Izzard\,\orcidlink{0000-0003-0378-4843}$^{1}$}
\author{David D. Hendriks\,\orcidlink{0000-0002-8717-6046}$^{1}$}

\affiliation{$^1$University of Surrey,  Stag Hill, University Campus, Guildford, GU2 7XH, United Kingdom}

\email{Corresponding author: n.rees@surrey.ac.uk}

\begin{abstract}

Mass changes due to strong stellar winds and binary mass transfer have a dramatic impact on the consequent evolution of stars.
This is generally not accounted for in population synthesis codes which are built using single star evolution models from full stellar evolution codes.
We produce a new grid of models using the 1D stellar evolution code \textit{MESA} which includes models with a range of core mass fractions, at each total stellar mass, evolved from core hydrogen exhaustion to the onset of core helium burning.
The model grid is used to produce an interpolation table, designed for use in population synthesis codes such as \textit{binary\_c}.
We test the interpolation table with a simple integration algorithm to evaluate its capability of reproducing accurate evolutionary tracks.
We test our method for stellar masses in the range $M=1-16~\Msun$ and show that it successfully reproduces the Hertzpsrung-gap and giant branch lifetime, the core mass at helium ignition and the HR diagram.

\end{abstract}

\maketitle

\section{Introduction}

Not all stars are the same as our most familiar star, the Sun.
Stars can be as low mass as $\approx0.08~\Msun$ and as high mass as $\gtrsim200~\Msun$ \citep{oeyStatisticalConfirmationStellar2005, vinkWindModellingVery2011, crowtherBirthLifeDeath2012}.
Knowing the mass of a star is paramount to predicting its evolutionary trajectory, with the most obvious example being the vast diversity in the final stellar remnants from white dwarfs to black holes.
However, the mass of a star is subject to change.
Mass-loss via stellar winds can have a dramatic effect on massive stars which experience strong stellar winds \citep{chiosiEvolutionMassiveStars1986, meynetGridsMassiveStars1994, 
vinkMasslossPredictionsStars2001,
eldridgeProgenitorsCorecollapseSupernovae2004, vanbeverenWolfRayetPopulationPredicted2007, vinkMassLossEvolution2008,
georgyYellowSupergiantsSupernova2012,
smithMassLossIts2014, 
renzoSystematicSurveyEffects2017, vinkRedSupergiantWind2023}.
Even low- to intermediate- mass stars ($0.8\lesssim M~/\Msun\lesssim10$) experience strong stellar winds when they reach the Asymptotic Giant Branch \citep[AGB,][]{bedijnPulsationMassLoss1988, vassiliadisEvolutionLowIntermediateMass1993, hofnerMassLossStars2018}.
However, stellar winds are not the only mechanism by which a star can change its mass.
Stars are commonly born within binary or multiple star systems.
At low mass ($0.8-1.2~\Msun$), approximately 40\% of stars are found within binary or multiple systems \citep{moeMindYourPs2017}, and the proportion only increases when higher masses are considered.
The majority of young massive stars are found in close binary systems, leading to binary interactions between the two stellar components \citep{podsiadlowskiPresupernovaEvolutionMassive1992, kobulnickyNewLookBinary2007, sanaBinaryInteractionDominates2012, sanaMultiplicityMassiveStars2013}.
Such binary interactions have dramatic consequences for the evolution and final fates of these stars.
Indeed, almost three quarters of all stars with $M\gtrsim 15~\Msun$ are strongly affected by binary interactions before exploding as a supernova
\citep{sanaBinaryInteractionDominates2012}.
An important binary interaction is mass transfer, which if unstable, can lead to stellar mergers \citep{podsiadlowskiPresupernovaEvolutionMassive1992, sanaBinaryInteractionDominates2012, deminkIncidenceStellarMergers2014, hennecoContactTracingBinary2023}.
Even in the case of stable mass transfer, the mass of both stellar components is significantly changed, thus altering the consequent evolutionary path.
For a review of the impact of companions on stellar evolution see \citet{demarcoDawesReviewImpact2017}.

Binary interactions and stellar winds change the total mass of a star.
Provided there is a strong enough composition gradient in the stellar interior from core-hydrogen burning, the star cannot adjust by adapting its core structure to match the new total mass \citep{neoEffectRapidMass1977,
braunEffectsAccretionMassive1995}.
Thus, if a star accretes mass from its companion, it has an under-massive core compared to constant mass evolution (CME). 
Similarly, if a star transfers mass to its companion or it loses mass in a stellar wind, it has an over-massive core compared to CME.
In cases of strong binary mass transfer, the star becomes a stripped star which has only a thin hydrogen-rich envelope surrounding the helium-rich core \citep{laplaceDifferentCorePresupernova2021, 
arancibia-rojasMassRangeHot2023,
duttaEvolutionaryNaturePuffedup2023, 
farmerNucleosynthesisBinarystrippedStars2023,
gotbergStellarPropertiesObserved2023}. 
Stellar evolution codes which solve the interior physics of stars compute the evolution of a star including mass changes.
For example, the stellar evolution code Modules for Experiments in Stellar Astrophysics \citep[\textit{MESA},][]{paxtonModulesExperimentsStellar2011,paxtonModulesExperimentsStellar2013,paxtonModulesExperimentsStellar2015,paxtonModulesExperimentsStellar2018,paxtonModulesExperimentsStellar2019a,jermynModulesExperimentsStellar2022} has been used to understand the evolutionary impacts of binary mass transfer and stellar winds \citep[e.g,][]{renzoSystematicSurveyEffects2017, laplaceDifferentCorePresupernova2021,bellingerPotentialAsteroseismologyResolve2023,hennecoContactTracingBinary2023, renzoRejuvenatedAccretorsHave2023,waggAsteroseismicImprintsMass2024}.
However, such codes are computationally expensive and thus cannot be used to model large populations of stars.
For such purposes, we can instead use population synthesis codes \citep{izzardPopulationSynthesisBinary2018}.
The basis of these population synthesis codes are built on the results from full stellar evolution codes. 
Historically, population synthesis codes have used analytical prescriptions derived from stellar evolution models \citep[e.g.][]{eggletonDistributionVisualBinaries1989, toutZeroageMainseqenceRadii1996, toutRapidBinaryStar1997}.
The future of population synthesis codes is pointed towards the use of interpolation tables to replace analytical prescriptions.
The motivation behind this is to reduce the loss of accuracy that occurs when a synthetic model is used over a full stellar evolution code.
The Method of Interpolation for Single Star Evolution (\textit{METISSE}) synthetic code \citep{agrawalFatesMassiveStars2020, agrawalModellingStellarEvolution2023} was recently developed to interpolate from stellar tracks produced by a variety of stellar evolution codes.
The method of interpolation has also been employed by the \textit{SEVN} \citep{speraMassSpectrumCompact2015,speraVeryMassiveStars2017} and \textit{COMBINE} \citep{kruckowProgenitorsGravitationalWave2018} codes to study the properties of gravitational wave progenitors.
Similarly, we aim to update the \textit{binary\_c} code \citep{izzardNewSyntheticModel2004, izzardPopulationNucleosynthesisSingle2006, izzardPopulationSynthesisBinary2009, izzardBinaryStarsGalactic2018, hendriksBinary_cpythonPythonbasedStellar2023,
izzardCircumbinaryDiscsStellar2023} with interpolation tables produced from grids of \textit{MESA} evolutionary tracks.
This is known as the Multi-Object Interpolation (\textit{MINT}) library and was introduced in \citet{mirouhDetailedEquilibriumDynamical2023}, in which a main-sequence interpolation table was used to study the impact of tides on open clusters.

The evolutionary grid we produce in this work notably differs from those previously computed by to the inclusion of stellar models with a range of core mass fractions, from stellar structures with massive envelopes to completely stripped helium stars.
This added dimension means that stellar properties can be estimated as a function of the core mass, envelope mass and a `time-proxy' which denotes the progression of a star through its current evolutionary phase.
This is a large step towards the development of a population synthesis code that self-consistently includes the impacts of mass changes due to binary mass transfer and strong stellar winds.
In addition, we save the temperature and density profiles of stars in the model grid to allow for the future development of the \textit{binary\_c} nucleosynthesis framework.
In this paper we focus on the evolutionary phase from core hydrogen exhaustion to core helium ignition.
This includes the Hertzpsrung-Gap (HG) and Red Giant Branch (RGB) phases.
Future work will extend this to later evolutionary phases.

In section \ref{sec:StellarModelling} we discuss the important considerations for stellar modelling before explaining the method for producing our model grid in section \ref{sec:Method}.
In section \ref{sec:Results} we present some interesting results from our model grid followed by a discussion of how our grid can be converted into an interpolation table for population synthesis in section \ref{sec:Discussion}
Finally we detail the limitations of our method in section \ref{sec:Limitations} and present final conclusions in section \ref{sec:Conclusions}.

\section{Stellar modelling}
\label{sec:StellarModelling}

For all computations we use version 23.05.1 of the 1D stellar-evolution code MESA.
Full MESA controls, including inlists and \texttt{run\_star\_extras} code, will be available on the MESA marketplace \footnote{\url{https://cococubed.com/mesa_market/inlists.html}}.
Computing a large grid of models over a range of masses without convergence failures can be a tricky task.
In Appendix \ref{append:convergence} we explain some of the methods used to reduce computational difficulties in our MESA models.
All stellar models in this work are computed at approximately Solar metallicity, $Z=0.02$.
This will be extended to a range of other metallicities in future work.

\subsection{Treatment of convection}
\label{sec:convection}

The treatment of convection is a large uncertainty in stellar modeling \citep{salarisConvectionStellarEvolution2007, stancliffePhysicsBonnetStellar2015}.
As in most 1D hydrostatic codes, convection is implemented in MESA using mixing length theory \citep[MLT,][]{bohm-vitenseUberWasserstoffkonvektionszoneSternen1958} with the formalism of \citet{henyeyStudiesStellarEvolution1965}.
The MLT method was recently reviewed by \citet{joyceReviewMixingLength2023}.
For the mixing length parameter, we use the Solar calibration, $\alpha_\mathrm{mlt}=1.931$, from \citet[]{cinquegranaSolarCalibrationConvective2022}.
As standard, convective boundaries in MESA are located based on sign changes in $y = \nabla_\mathrm{rad}-\nabla_\mathrm{ad}$ (or $y = \nabla_\mathrm{rad}-\nabla_\mathrm{L}$ for the Ledoux criterion), where $\nabla_\mathrm{rad}$ and $\nabla_\mathrm{ad}$ are the adiabatic and radiative temperature gradients respectively and $\nabla_\mathrm{L}$ is the Ledoux temperature gradient \citep[Eq 11]{paxtonModulesExperimentsStellar2013}.
% if need reference for gradients use (Kippenhahn & Weigert 1990).
However, it has long been known that the mixing of material in a star extends slightly beyond the region predicted by the Schwarzschild or Ledoux stability criteria. 
This occurs because the convective eddies have inertia, and they “overshoot” into adjacent stable layers before dissipating.
The consequence of this is the increased size of convective regions compared to that calculated using the Schwarzschild or Ledoux criterion.
More recent versions of MESA include a new prescription for calculating convective boundaries, known as convective premixing \citep[][]{paxtonModulesExperimentsStellar2019a}. 
The convective premixing mixing algorithm considers whether radiative cells on the outside of the convective boundary would become convective if they are mixed completely with the rest of the convective region.
This scheme notably increases the mass of convective cores during core hydrogen and core helium burning, such that they are more consistent with observational constraints.
Extra mixing beyond convective boundaries can also be implemented with convective overshooting.
At the bottom of the convective envelope we force the diffusion coefficient to decrease exponentially with distance from the convective boundary by using exponential overshooting with efficiency parameter $f_\mathrm{ov} = 0.0174$, which is the Solar calibration from \citet{choiMESAISOCHRONESSTELLAR2016}.
Overshooting is helpful to prevent splitting of the convective envelope on the giant branch which can lead to spurious results for first dredge-up and the red giant branch bump.

The size of the convective core on the main sequence (MS) influences the stellar lifetime as well as later evolutionary stages via the resulting mass of the H-exhausted core.
There is a plethora of observational evidence suggesting that the size of the convective core is greater than that given by the linear-stability criteria of Schwarzschild and Ledoux \citep{maederExtentMixingStellar1981, demarqueY2IsochronesImproved2004, aertsAsteroseismologyBinaryStars2013, claretDependenceConvectiveCore2016, andersConvectiveBoundaryMixing2023}.
Thus, some extension of the core convection zone is required.
On the MS we implement an approximation of convective penetration which occurs when motions mix the entropy gradient towards the adiabatic in a region that is stable by the Schwarzschild criterion \citep{roxburghOriginSupergranulationGiant1979,andersStellarConvectivePenetration2022,andrassySelfconsistentModelConvective2023}.
This assumes that mixing is fast enough to change the thermal stratification beyond the Schwarzschild boundary.
It can be implemented using step overshooting, which forces uniform mixing up to a fixed distance beyond the convective boundary.
The radial extent of this `convective penetration layer' is set by the free parameter in step overshooting, $\alpha_\mathrm{ov}$.
For the value of $\alpha_\mathrm{ov}$ we use the mass-dependent fitting formulae of \citet[Eqs. 11-16]{jermynConvectivePenetrationEarlyType2022} which are based on the simulations and theory of \citet{andersStellarConvectivePenetration2022}.
For masses greater than the range computed in their work ($1.1-60~\Msun$) we extrapolate the same formulae due to the lack of a more informed choice.
However, the step overshoot prescription available as an inlist option in MESA only mixes the composition in the overshoot layer whilst the temperature gradient is kept radiative.
We use the routine provided in \citet{andrassySelfconsistentModelConvective2023} to force the temperature gradient to equal the adiabatic temperature gradient in this region.
During core-hydrogen burning we also use the convective premixing scheme to be consistent with \citet{jermynConvectivePenetrationEarlyType2022} and \citet{andrassySelfconsistentModelConvective2023}.

\subsection{Numerical controls for convergence}
\label{append:convergence}

We highlight some of the \textit{MESA} controls that we have found useful to aid convergence in our stellar grid.

Firstly, to produce pre-main sequence models over a range of initial masses, we create starting models with uniform composition, a core temperature below that required for nuclear burning and uniform contraction such that the luminosity is high enough for the structure to be fully convective.
In \textit{MESA} this can be achieved using the \texttt{create\_pre\_main\_sequence\_model} function.
We find that using a high starting temperature (\texttt{pre\_ms\_T\_c}) of $T_\mathrm{c}=9\times10^5~\mathrm{K}$ works best for starting models that reliably converge to a pre-main sequence structure. 

Computation of massive stars ($\Mi \gtrsim 15~\Msun$) within 1D stellar evolution codes is tricky due to their numerically (and probably physically) unstable envelopes \citep{maederChangesSurfaceChemistry1987, paxtonModulesExperimentsStellar2013}.
Iron opacity bumps in the envelope at $\log (T/\mathrm{K}) \approx 5.3$ and $6.3$ cause the local radiation pressure to dominate and the luminosity approaches the Eddington luminosity.
This can lead to inversions in density and gas pressure which cause computational difficulties and result in prohibitively short timesteps.
The higher the metallicity, the greater the problem due to the increase in strength of the iron opacity bumps \citep[Fig.~38]{paxtonModulesExperimentsStellar2013}.
These unstable regions can be tackled by using techniques to reduce the superadiabatic gradients that arises in radiation-dominated convective regions.
\citet{paxtonModulesExperimentsStellar2013} developed a treatment of convection called MLT++ for this purpose.
This allows uninterrupted evolution from ZAMS to core collapse although the results of 1D stellar evolution calculations for the late evolutionary phases of massive stars will be highly uncertain.
\citet{jermynModulesExperimentsStellar2022} discuss that a limitation of the MLT++ treatment is that it is a non-local, explicit method which can lead to large step-to-step variations, unphysical results and solver issues.
Thus, they implemented a fully implicit and local alternative to MLT++ (\texttt{use\_superad\_reduction}) allowing for the modeling of a larger range of masses and metallicities.
We use it for all evolutionary phases for viable evolution of massive models, although, it is phased in at the start of the main sequence to prevent it from causing solver issues when converging a pre-main sequence model.

At the end of core-hydrogen burning the convective core boundary can become unstable leading to a sudden expansion of the convective core which is otherwise retreating. 
This leads to strong MS rejuvenation and can cause major convergence issues which result in computation failure. 
To prevent this, and allow for the production of TAMS models which can be used for the next phase, we phase out the convective core overshooting by linearly reducing the efficiency parameter after the core boundary has formed ($\Xc\lesssim 0.1$).
Once $\Xc<0.01$, convective overshooting and convective premixing are turned of entirely.

An additional problem occurs in the envelopes of massive models with $M\gtrsim 15~\Msun$ on the HG.
During this phase a large portion of the envelope is convective, although the outer envelope is radiative.
The bottom of the convective region can erroneously move past the H-exhausted core boundary and dredge-up material.
This is not the same as the first dredge-up because the convective region is not connected to the surface and the extension of the region occurs suddenly rather than as a gradual growth over time.
This causes sudden and random changes in the core mass as well as convergence problems which can terminate the model.
As this is most likely unphysical, we want to prevent this from occurring to keep models running and give smooth core mass evolution.
We prevent mixing from occurring inside the hydrogen burning shell by turning of composition changes due to mixing for mass shells in the region with $10^{-8}\lesssim X \lesssim 10^{-2}$.
This routine is used only for masses $M>15~\Msun$ so as to not impact first dredge-up.

\section{Method}
\label{sec:Method}

\begin{figure*}
 \centering
\includegraphics[width=\linewidth,height=0.9\textheight,keepaspectratio]{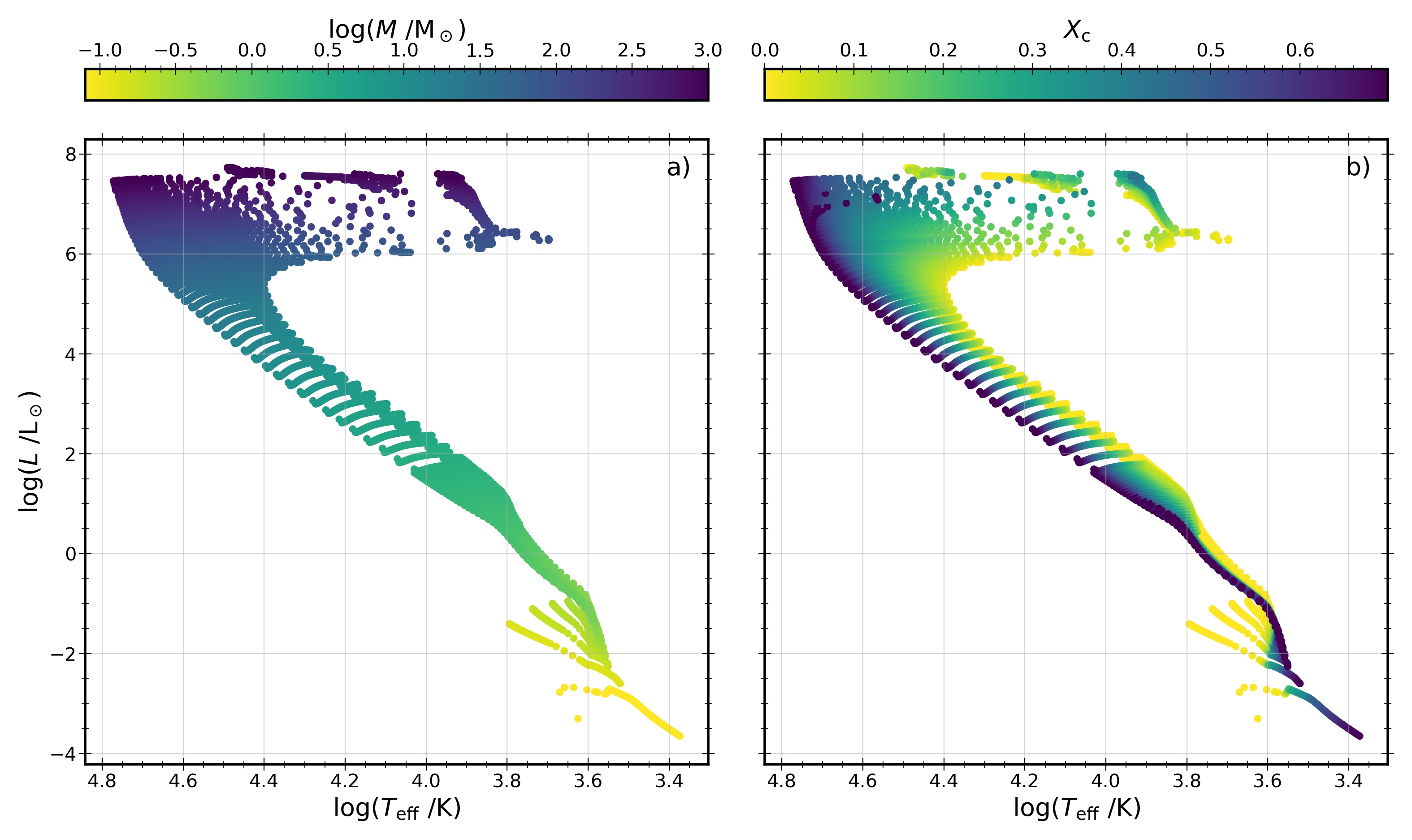}
    \caption{Main sequence evolution of models with $0.08\leq M~/\Msun \leq 1000$, coloured by their mass (panel a) and central hydrogen mass fraction (panel b).}
    \label{fig:MS_HR}
\end{figure*}

\begin{figure}
 \centering
\includegraphics[width=\linewidth,height=0.9\textheight,keepaspectratio]{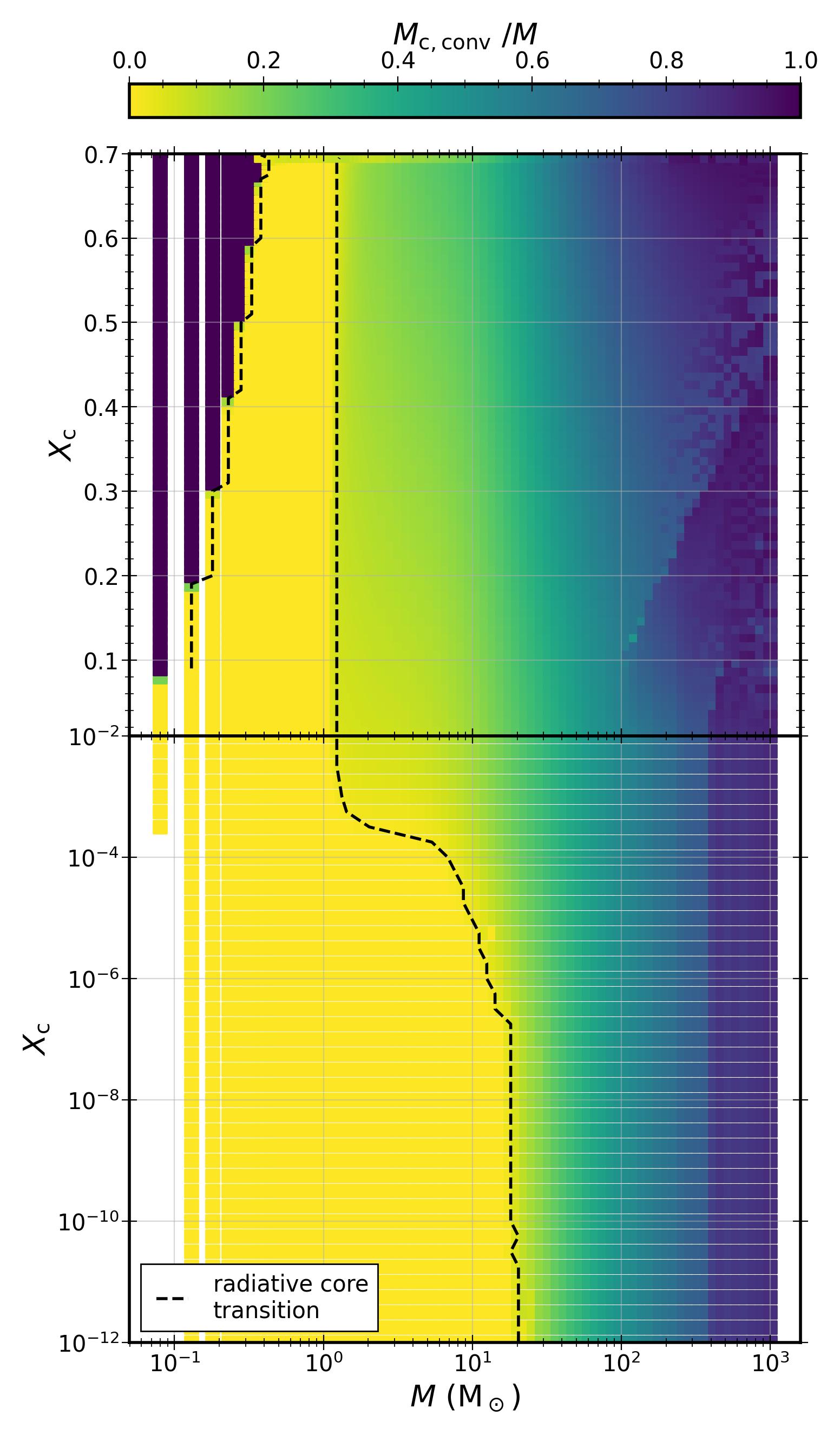}
    \caption{The mass of the convective core as a fraction of the total stellar mass, $M_\mathrm{c,conv}/M$, on the MS as a function of the stellar mass, $M$, and central hydrogen mass fraction, $\Xc$. Yellow areas indicate radiative core-hydrogen burning whilst dark purple areas indicate fully-convective models. See also \citep[][Fig. 1]{mirouhDetailedEquilibriumDynamical2023}.}
    \label{fig:MS_conv_core_mass}
\end{figure}

We start by computing main-sequence (MS) models without mass-loss in the range $0.08\leq \Mi~/\Msun \leq 1000$.
All masses are evolved from the pre-main sequence (PMS) to core-hydrogen exhaustion when the central-hydrogen mass fraction $\Xc < 10^{-12}$.
In Fig.~\ref{fig:MS_HR} we present the Hertzsprung-Russell (HR) diagrams of all our models labelled by stellar mass (\textit{a}) and central hydrogen mass fraction (\textit{b)}.
Some very-low-mass stars evolve leftwards on the HR diagram towards the WD cooling track before exhausting hydrogen in the core, although the MS lifetimes of these stars is longer than the age of the Universe.
In Fig.~\ref{fig:MS_conv_core_mass} we plot the convective core mass fraction as a function of $M$ and $\Xc$ in all our MS models.
Our models with $\Mi\lesssim0.4~\Msun$ are fully convective for some portion of the MS which enriches the envelope in helium.
Radiative core-hydrogen burning occurs in our models with $0.4 \lesssim \Mi~/\Msun \lesssim 1.2$.
Above this, the models switch to convective core-hydrogen burning.
The mass of the convective core increases relative to the stellar mass as the stellar mass increases.
The helium-core boundary is defined to be the outermost location where the hydrogen mass fraction, $X < 0.01$ and the helium mass fraction $Y > 0.1$.
Thus, in all our models there is a non-zero He-core mass, $\Mc$, when $X_\mathrm{c}\leq0.01$.

At the end of the MS, the total stellar mass is altered, by adding or removing mass from the stellar surface, to produce a range of models with different core mass fractions.
This process must be done after a He-core has formed but before the star evolves across the Hertzsprung-Gap (HG) which occurs out of thermal equilibrium (section~\ref{sec:HGThermalEquilibrium}).
Thus, we save models when $\Xc \approx 10^{-4}$ which we define to be the terminal-age main sequence (TAMS).
These TAMS models are used as starting models for the mass change process.
TAMS models from fully-convective MS stars ($\Mi\lesssim0.4~\Msun$) are not used as they have helium-rich envelopes which affects the consequent evolution.
For example, the hydrogen burning shell processes material that has a higher helium content and thus the core mass grows more rapidly leading to less degenerate cores at helium ignition.
In any case, these stars have very long MS lifetimes and thus rarely contribute to observed populations of post-MS stars.
We describe the mass change process in detail in Section.~\ref{sec:masschange}.

After the mass change is complete, the models are evolved until helium ignites in the core.
We terminate due to helium ignition when the central helium abundance, $Y_\mathrm{c}$, drops by $> 0.01$ from its maximum value of $\approx1-Z$.
We also terminate models if the degeneracy at the centre exceeds a limit, $\etac>50$, where $\eta$ is the electron chemical potential in units of the Boltzmann constant multiplied by the temperature.
This terminates the computation of low-mass models with $M\lesssim0.32~\Msun$, that cannot grow massive enough cores to ignite helium.

To provide sufficient resolution in the TAMS core mass, we must run the MS model grid with a high resolution in the initial mass.
At low masses, in the range $0.08\leq\Mi~/\Msun \leq 2.6$, we use mass steps of $0.03~\Msun$ to produce core masses which resolve the RGB phase transition at the switch between degenerate and non-degenerate helium ignition \citep{sweigartDevelopmentRedGiant1990, cassisiRedGiantBranch2016}.
Above this, $2.6<\Mi~/\Msun\leq1000$, we compute the MS for an additional one-hundred initial stellar masses, equally spaced in logarithmic space.
The total stellar masses produced by the mass change process are spaced such that there is an increase of $\approx10\%$ between consecutive masses.
The exact values are chosen such that they are all equal to an initial mass included in the MS grid.
This ensures there are constant mass evolution (CME) models which do not undergo the mass change process.
We can use these to validate that the mass change process does not lead to any undesired effects.
Our grid includes under- and over-massive core models at all total masses with the exception of low mass stars ($M\lesssim 1.2~\Msun$) for which the CME models give the lowest core mass possible from radiative core-hydrogen burning.

\subsection{Changing the stellar mass}
\label{sec:masschange}

We must alter the stellar mass by adding or removing mass from the stellar surface without impacting the core structure.
Thus, we turn off composition changes due to nuclear burning by setting the \textit{MESA} control \texttt{dxdt\_nuc\_factor} to zero. 
We also prevent composition changes by convective mixing as this can lead to undesired composition profiles produced by temporary convective regions during the mass change process.
This can be done by setting the control \texttt{mix\_factor} to zero.
These controls allow the models to be frozen in time at the TAMS whilst the total mass is changed.
If the target total mass is greater than the initial stellar mass, we accrete mass with the same composition as the surface at a constant rate. 
The magnitude of the accretion rate does not make a difference to the stellar structure which is frozen in time, and thus is chosen purely to minimize the computation time.
We initially try a mass accretion rate of $10^{-8}~\Msun~\mathrm{yr}^{-1}$ but if the mass accretion is too slow we instead increase it to  $10^{-4}~\Msun~\mathrm{yr}^{-1}$.
This higher mass accretion rate is required for the higher mass models which are constrained to shorter time steps due to computational difficulties relating to the Eddington limit (Appendix~\ref{append:convergence}).
If the target total mass is less than the initial stellar mass, we strip the envelope with a constant mass loss rate of $10^{-8}~\Msun~\mathrm{yr}^{-1}$.
However, this causes convergence problems when the model is stripped to a small layer of envelope on top of the helium core.
This occurs for reasons similar to the Fe-peak instability found in TP-AGB and post-AGB stars \citep[][]{lauEndSuperAGB2012, reesEvolvingInstabilitiesThermally2024}.
Instead, when the target total mass is close to the helium core mass ($<1.25~\Mc$), we find it computationally smoother to remove all mass shells outside of the chosen boundary in one step and then takes multiple timesteps to relax the composition and entropy of the model back to an equilibrium solution.
This can be done using the \textit{MESA} routine \texttt{star\_relax\_to\_star\_cut}.
We find this method to be more successful than a wind-like mass loss for stripping models down to thin envelopes.
Note that because the model composition is frozen, it does not matter how we get to the final core mass fraction and thus accretion and mass-loss rates are chosen purely for numerical convenience.
Once the target total mass has been achieved, by mass loss, accretion or cutting, the model is evolved through a relaxation phase lasting ten Kelvin-Helmholtz timescales to regain thermal equilibrium. 
These models can then be used to resume the evolution following core-hydrogen exhaustion.

\subsection{Interpolation variables}
\label{sec:InterpVariables}

\begin{figure}
 \centering
\includegraphics[width=\linewidth,height=0.9\textheight,keepaspectratio]{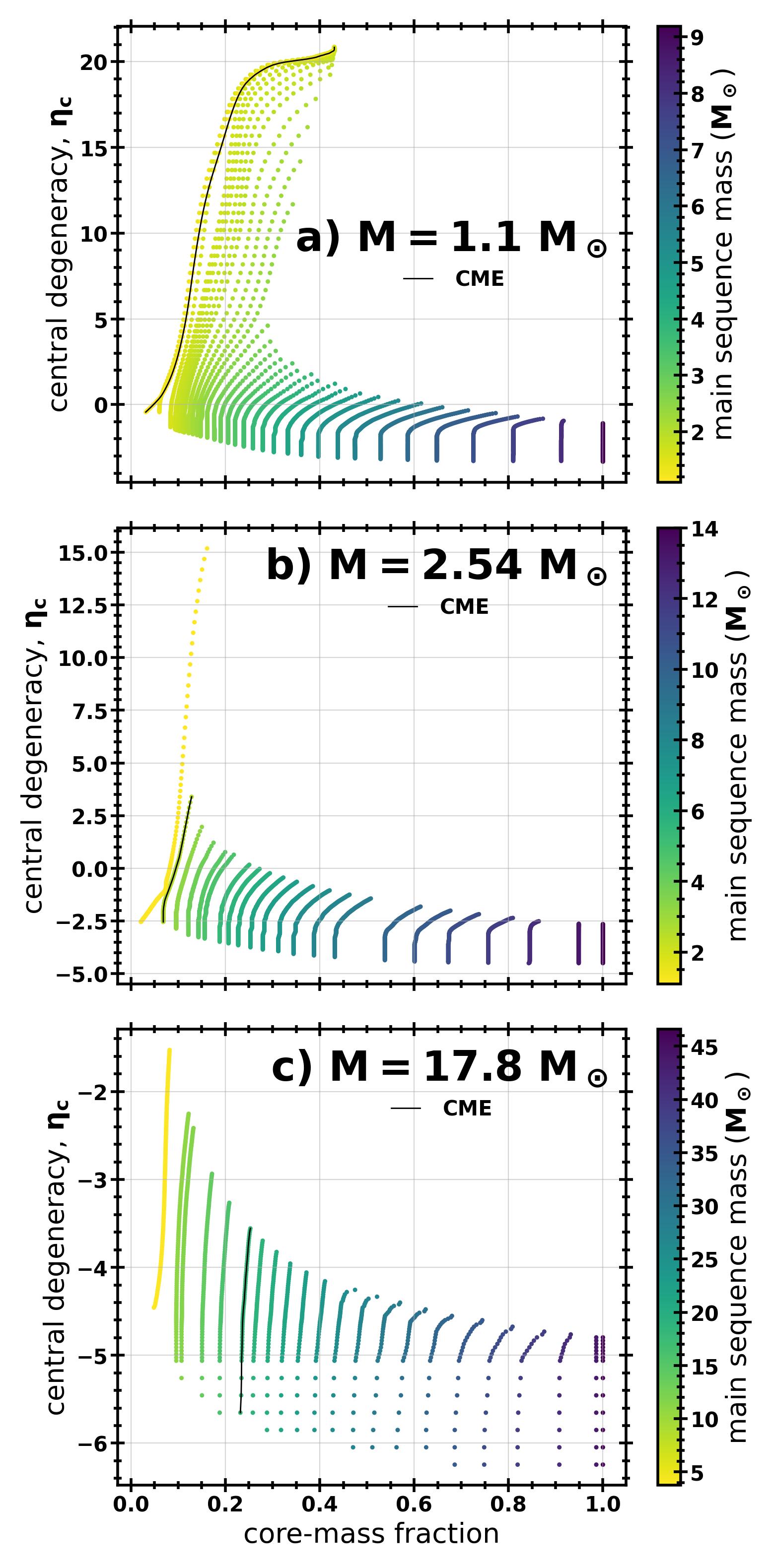}
    \caption{Evolution in the core mass fraction- central degeneracy parameter space of stars with masses $1.1$ (\textit{a}), $2.54$ (\textit{b}) and $17.8~\Msun$ (\textit{c}) and a range of core mass fractions. The main sequence mass used to compute each core mass is labelled by the colour. Also plotted in solid black is the constant mass evolution (CME) model for each stellar mass.}
    \label{fig:GB_tracks}
\end{figure}

To produce interpolation tables from the evolutionary tracks we must first decide our `interpolation variables' which define the current state of the star.
Other quantities, such as the luminosity, can then be interpolated as a function of these state variables.
The most obvious variable to use is the total stellar mass and thus we use this as our first interpolation variable for all phases of evolution.
The total stellar mass also has the advantage that it is easily kept constant, by neglecting mass loss, and thus can be set to chosen values.
Our grid contains models with a range of core and envelope masses and thus we use the core mass fraction as our second interpolation variable.
For the third interpolation variable, we require a quantity that describes the progression of the stellar model throughout its current evolutionary phase.

In the first \textit{MINT} paper \citep{mirouhDetailedEquilibriumDynamical2023}, the central hydrogen mass fraction, $X_\mathrm{c}$, is used as a `time-proxy'.
This works because it is a monotonic quantity and thus can define the state of evolution through the MS.
Note we do not use time as an interpolation variable because time is not a useful co-ordinate when stars interact. 
For example, MS stars can become rejuvenated in binaries and appear younger than they are.
By using time-proxy interpolation variables that are innate properties of the stellar core, we can naturally deal with stellar mass changes.
This is in contrast to codes that use time as an interpolation variable, which then must invoke the concept of an `effective initial mass' when mass changes occur \citep[e.g.][]{hurleyEvolutionBinaryStars2002,agrawalModellingStellarEvolution2023}.

To find our `time-proxy' variable for the phase after core-hydrogen exhaustion and before core-helium ignition we look for a quantity that monotonically increases or decreases.
As our stellar models evolve off the MS, across the HG and up the first giant branch (FGB), the core contracts and cools, causing the degeneracy to increase.
In Fig.~\ref{fig:GB_tracks} we show the evolutionary tracks at three different total stellar masses which each include a number of models starting with different core mass fractions.
In subplot \textit{a}) we display a low-mass example, which includes both tracks that terminate by igniting helium degenerately, and those that don't due to starting with a more massive core at the TAMS.
In subplot \textit{c}) we display a high-mass example, in which all tracks start with sufficient core mass to ignite helium gently.
Finally, in subplot \textit{b}) we display a mass at the transition between having both degenerate and non-degenerate tracks and only having non-degenerate tracks.
We discuss helium ignition in detail in section \ref{sec:HeIgn}.
Model tracks which include a FGB phase show some core-mass growth via the hydrogen-burning shell.
Vertical tracks are seen for high-mass models that ignite helium before reaching the FGB and stripped stars which have very inefficient hydrogen-burning shells due to the lack of a hydrogen rich envelope.
During hydrogen-shell burning we thus can use the central degeneracy, $\etac$, as the time-proxy to uniquely define the state of the core as it transitions from core hydrogen exhaustion to core-helium burning.

In the case of strong mass accretion ($\Mc/M\lesssim0.05$ and $M\gtrsim2.5~\Msun$), we find there is an additional readjustment phase at the TAMS when nuclear burning is turned back on.
Instead of crossing the HG with increasing core degeneracy, there is a brief heating phase in which the core degeneracy drops.
The core mass then grows rapidly due to the increased temperatures.
This seems to point to a natural lower limit of stable stellar structures of $\Mc/M \gtrsim 0.05$ for $M\gtrsim 2.5~\Msun$.
With the exception of this readjustment period for strong mass accretion, the central degeneracy is monotonically increasing quantity for all stellar evolution tracks.
Thus, we can use it as our time-proxy interpolation variable.

\subsubsection{Time-proxy resolution}

\begin{figure}
 \centering
\includegraphics[width=\linewidth,height=0.9\textheight,keepaspectratio]{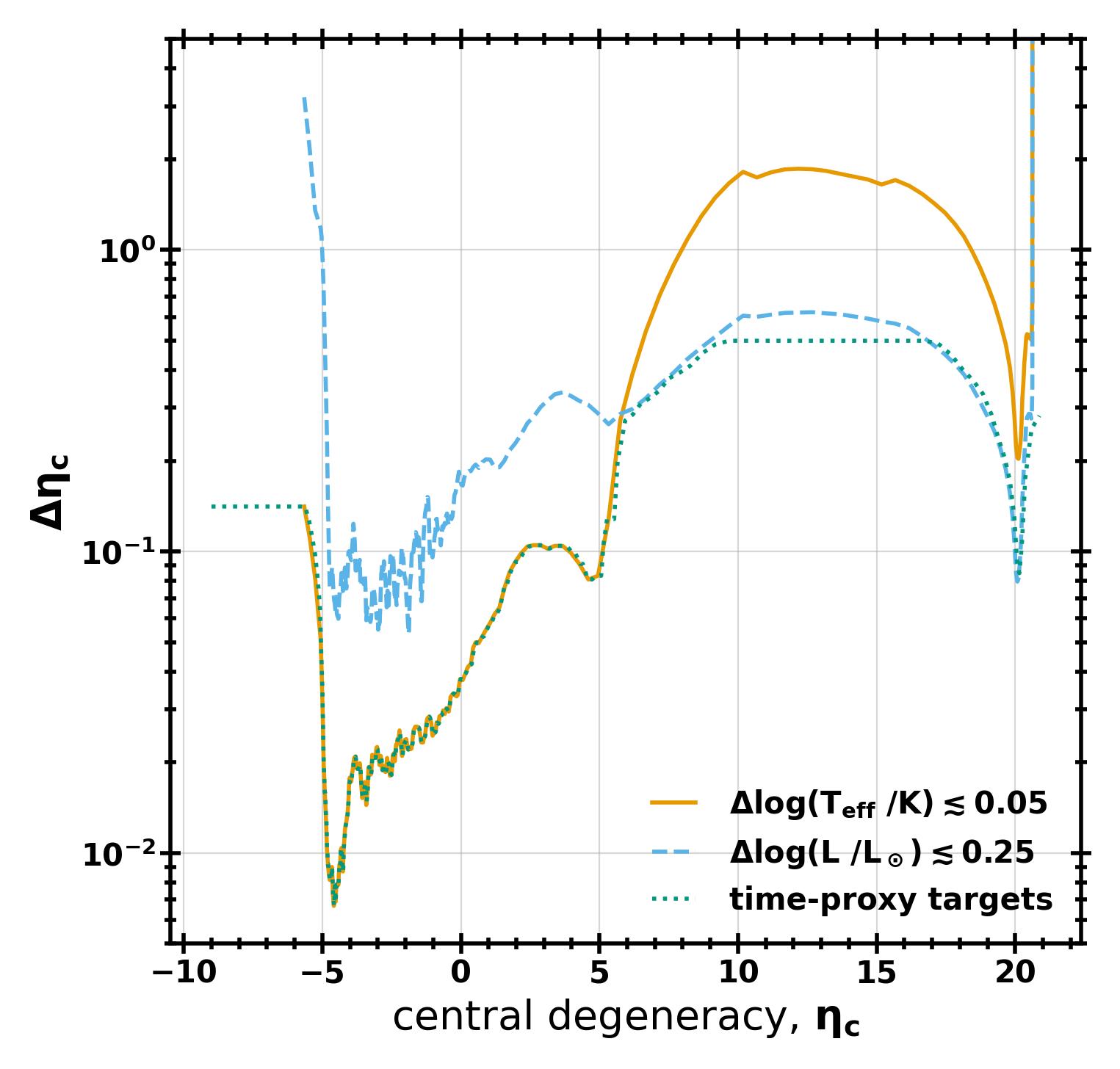}
    \caption{The required central degeneracy, $\etac$, resolution of our \textit{MESA} tracks to meet the maximum allowed change in the effective temperature (solid, orange) and the luminosity (dashed, blue) at constant mass and core mass fraction. Also plotted is the chosen time-proxy save target resolution to meet both criteria (dotted, green).}    
    \label{fig:degen_res}
\end{figure}

To provide an interpolation table which can reconstruct our evolutionary tracks we must save the model data with sufficient resolution in the time-proxy.
We set `save targets' for the time-proxy variable, $\etac$, which tell \textit{MESA} to save model data when $\etac$ reaches these values within some small tolerance.
Thus, the save targets set the resolution of the interpolation table in the time-proxy variable.
To ensure we retain sufficient resolution we use model tracks from an exploratory first iteration to compute the range of change of the stellar luminosity, $L$ and effective temperature, $\Teff$, with respect to $\etac$.
For each $\etac$ the maximum rate of change from all the model tracks is computed.
We then set the save targets such that changes in the logarithms of the stellar luminosity and radius between save targets are less than some tolerance for all model tracks.
In Fig.~\ref{fig:degen_res} we plot the required $\Delta\etac(\etac)$ when the allowed change in effective temperature is $\Delta\log_{10}(\Teff~/\mathrm{K}) \lesssim 0.05$ (solid line) and the allowed change in luminosity is $\Delta\log_{10}(L~/\Lsun) \lesssim 0.25$ (dashed line).
The time-proxy targets are then chosen to meet the minimum resolution required at every point (dotted).
We also cap the maximum allowed $\Delta \etac$ at $0.5$.
This results in a total of 421 time-proxy save targets.

We find that increased resolution is required at the helium-flash around $\etac \approx 20$, due to the stellar luminosity rapidly increasing at the RGB-tip.
High resolution is also required around $\etac\approx-5$ where the most massive stars that ascend the RGB rapidly move across the HG with a decreasing effective temperature. 
More massive stars, at lower degeneracy, ignite helium on the HG and thus there is limited change in the surface properties during this phase.
Increasing the time-proxy resolution doesn't strongly impact the computation time of models but does increase the amount of data saved and thus the final table size. 

\subsubsection{Core mass fraction resolution}

\begin{figure}
 \centering
\includegraphics[width=\linewidth,height=0.9\textheight,keepaspectratio]{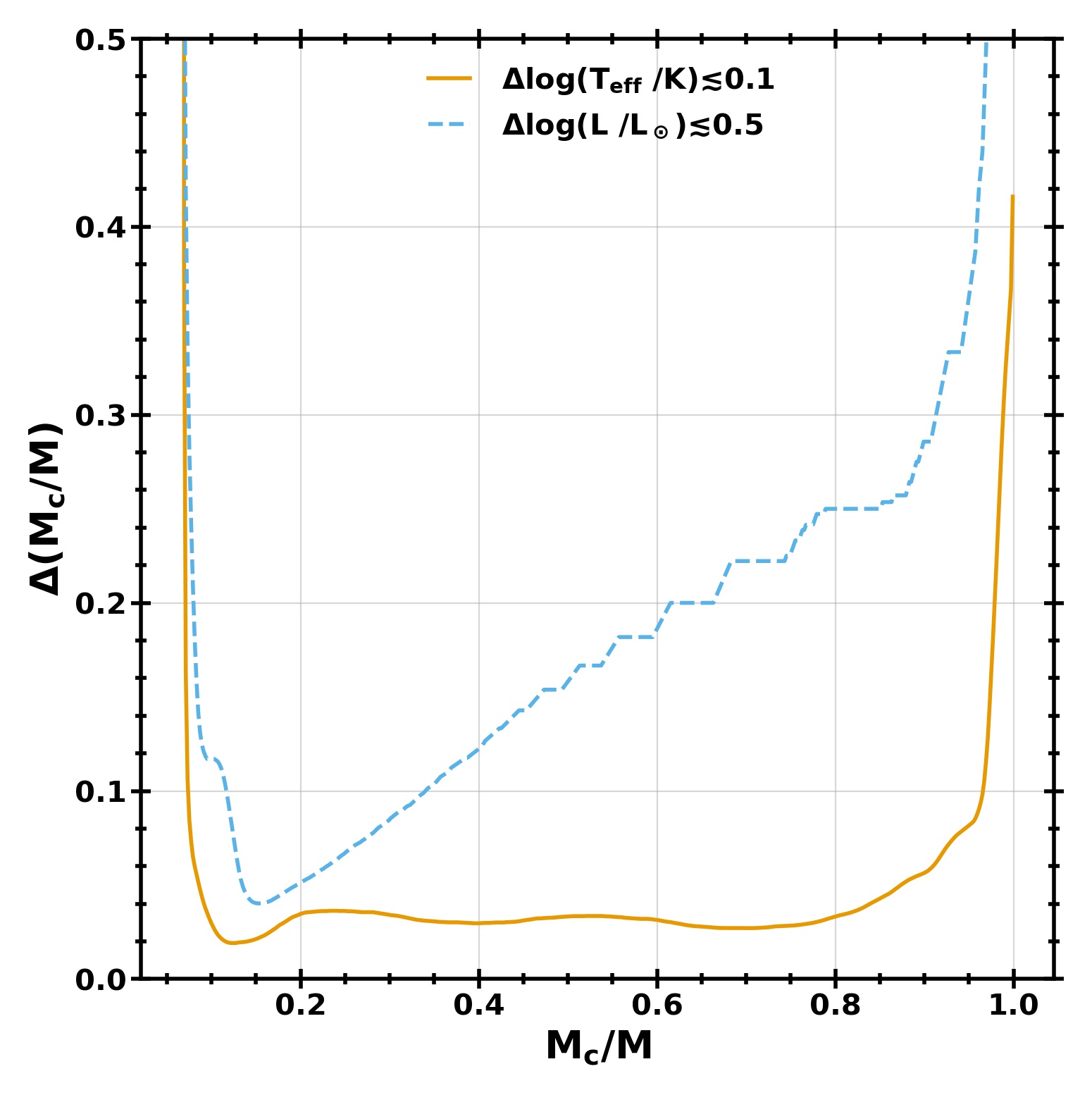}
    \caption{The core mass fraction resolution, $\Delta\Mc/M$, of our \textit{MESA} tracks required to meet the maximum allowed change in the effective temperature (solid orange) and the luminosity (dashed blue) at constant mass and central degeneracy.}
    \label{fig:Mc_res}
\end{figure}

The accuracy of interpolation also depends on the resolution of the core mass fraction variable.
At each total stellar mass, we run a number of \textit{MESA} tracks with different initial core masses to fill the time proxy-core mass parameter space (Fig.~\ref{fig:GB_tracks}).
If the change in core mass between consecutive tracks is too large, there are large gaps in the parameter space which lead to inaccurate interpolation.
Thus, we space out the initial core masses produced at the TAMS for each stellar mass. 
However, due to the growing core mass, our \textit{MESA} tracks are curved in the parameter space and thus the spacing between two adjacent tracks varies.
In addition, we do not know a priori how the stellar models will behave.
Thus, we instead use a run-time approach which checks the spacing between existing adjacent tracks and if it exceeds a target amount, an additional track is run to bisect the space.
We start by running the lowest core-mass model available, the constant mass evolution (CME) model and the fully-stripped model.
The spacing is then checked and the next models to run are chosen to target the space halfway between the existing models.
This process continues until the target resolution is met, or there are no further models to use.
It is thus important that we run the MS at sufficiently-high mass resolution to provide a good resolution of core masses available for the consequent phases.
For example, we run MS models with increased resolution around the RGB phase transition mass such that this can then be resolved in the GB tracks.

To decide the target resolution in the core mass fraction, we compute the rate of change of some quantity, such as the luminosity, with respect to the core mass fraction at constant values of the stellar mass and time-proxy. 
We can then calculate the allowed change in the core mass fraction, $\Delta \Mc/M$, such that the quantity changes by less than some critical value.
This is the same method used above to calculate the time-proxy resolution however now we are looking at the difference between \textit{MESA} tracks with different initial core masses.
At all unique combinations of the stellar mass and time-proxy we calculate $\Delta \Mc/M$ as a function of $\Mc/M$ and then take the minimum value so that the criteria is met everywhere in the parameter space.
We take the allowed $\Delta \Mc/M$ for the $0.05$ quantile to exclude extreme outliers.
This ensures that $\approx 95\%$ of the parameter space has sufficient resolution to meet the criteria.
In Fig.~\ref{fig:Mc_res} we plot the allowed $\Delta \Mc/M$ as a function of $\Mc/M$ such that the allowed change in effective temperature is $\Delta\log_{10}(\Teff~/\mathrm{K}) \lesssim 0.1$ (solid line) and the allowed change in luminosity is $\Delta\log_{10}(L~/\Lsun) \lesssim 0.5$ (dashed line).
The highest resolution is required around $\Mc/M\sim0.15$, i.e. models with large envelope masses.
Due to the difficulty of implementing a core mass fraction resolution as a function of $\Mc/M$ we use a constant target value of $\Delta \Mc/M=0.05$.
This means that a significant portion of the parameter space is better resolved than our criteria.

\begin{figure}
 \centering
\includegraphics[width=\linewidth,height=0.9\textheight,keepaspectratio]{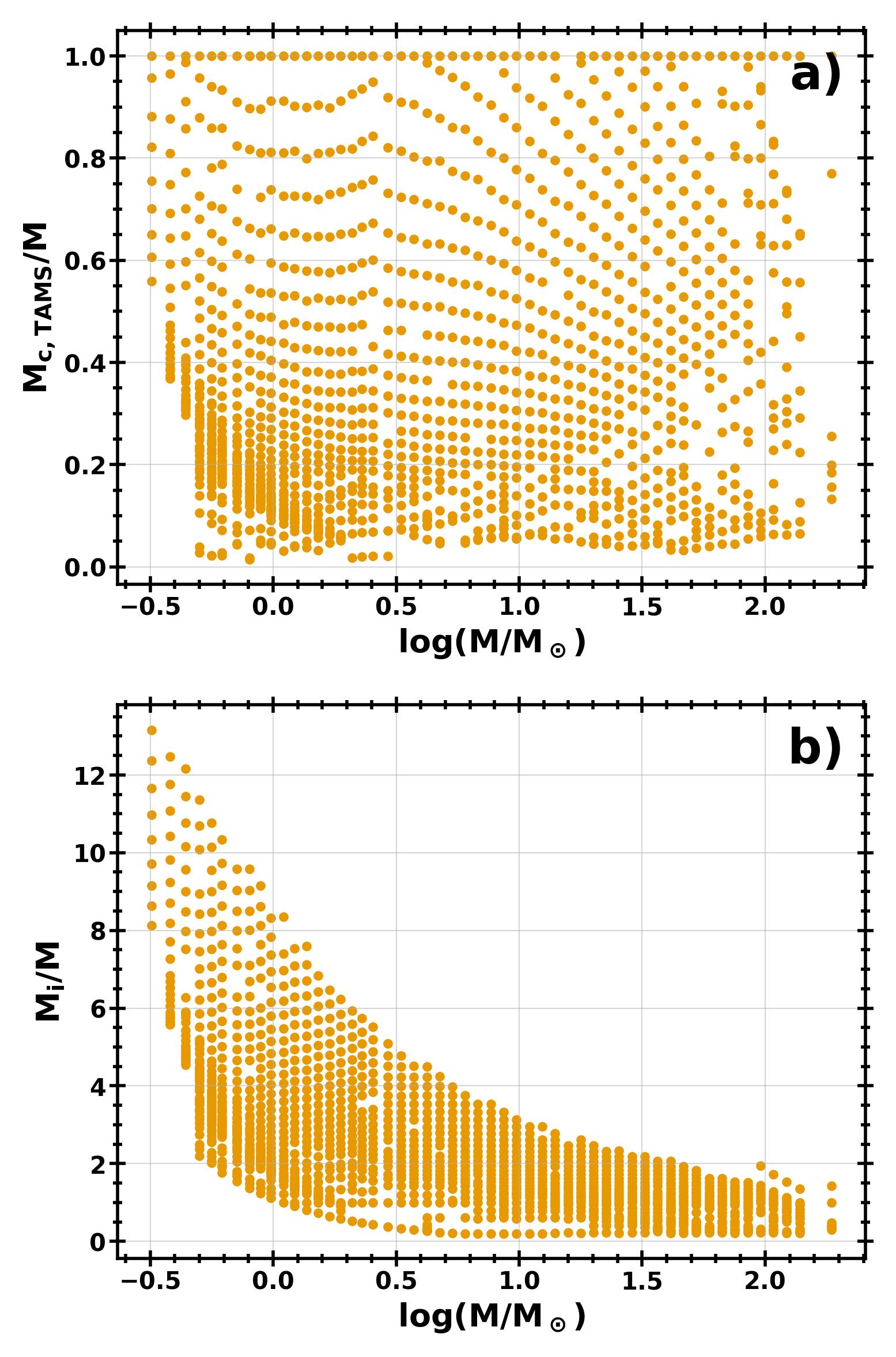}
    \caption{The total stellar mass, $M$, vs the a) TAMS core mass fraction, $\Mc/M$, and b) initial mass fraction, $\Mi/M$, for all the evolution tracks in the grid.}
    \label{fig:TAMS_models}
\end{figure}

To illustrate the parameter space covered by our method, we have plotted the a) TAMS core mass fraction, $\Mc/M$, and b) initial mass, $\Mi/M$ for all the evolution tracks in the grid in Fig.~\ref{fig:TAMS_models}.

\section{Results}
\label{sec:Results}

% Knowing the stellar luminosity, $L$, and effective temperature, $\Teff$ is vital for comparing theoretical results to observational studies.
% In addition, knowing the stellar radius is essential for predicting binary interactions during population synthesis calculations.
% We thus study the HR diagrams for a selection of tracks to get a feel for the underlying behaviour of the models.

\begin{figure}
 \centering
\includegraphics[width=\linewidth,height=0.9\textheight,keepaspectratio]{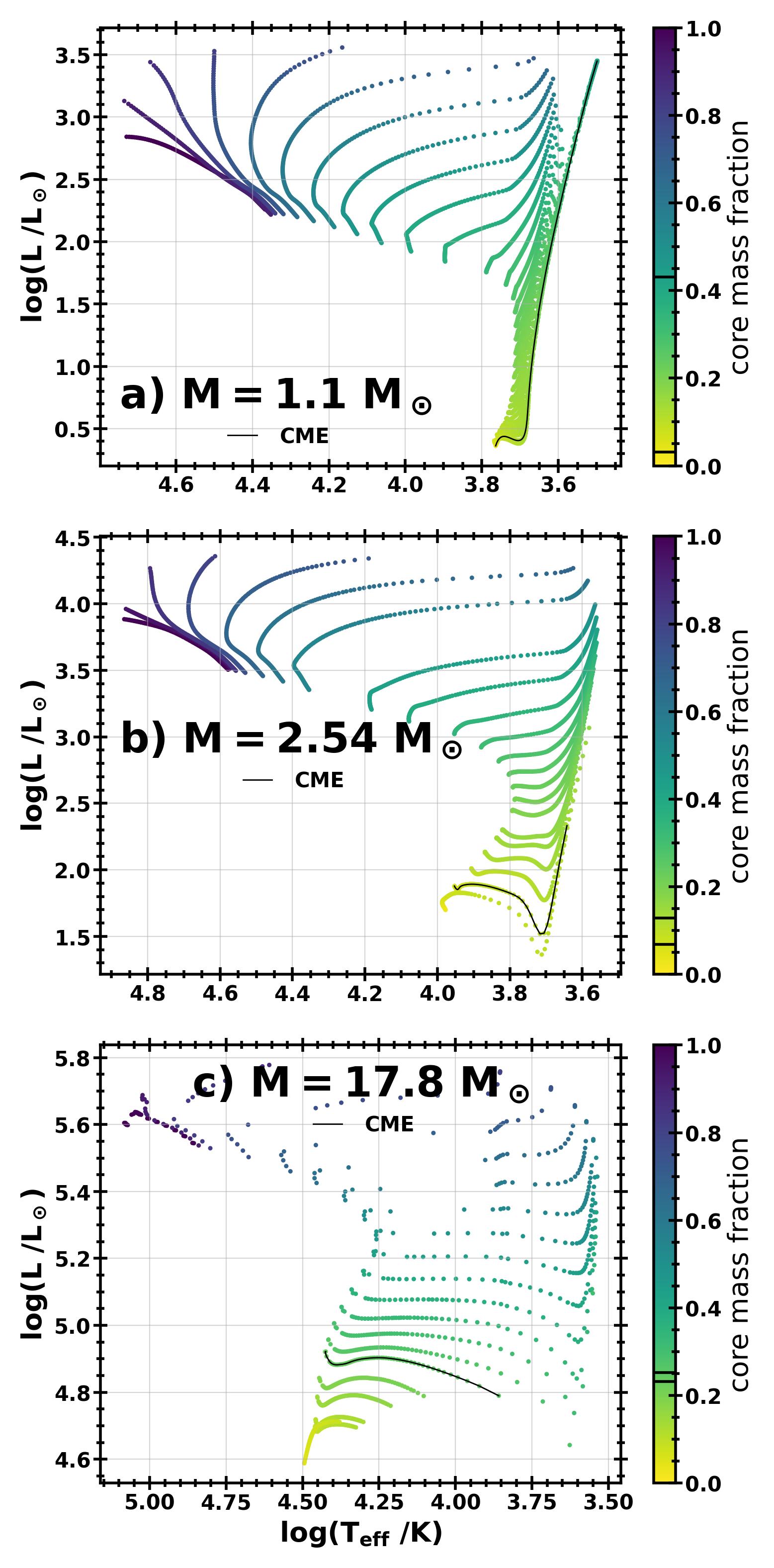}
    \caption{Hertzsprung-Russell (HR) diagrams of the hydrogen-shell burning phase, showing our tracks with a range of core mass fractions and total stellar masses of (a) $1.03~\Msun$, (b) $2.53~\Msun$ and (c) $18.1~\Msun$.
    Constant mass evolution (CME) models are highlighted with black lines.
    We also include two black lines on the color bars to indicate the starting and ending core mass fractions of the CME model tracks.}
    \label{fig:GB_HR}
\end{figure}

After hydrogen exhaustion in the core, hydrogen burning moves to a shell surrounding the core.
The He-rich core grows in mass until the temperature in the core, $T_\mathrm{c}$, is sufficient ($\approx 10^8~\mathrm{K}$) to ignite helium.
In Fig. \ref{fig:GB_HR} we show the HR diagram evolution for this phase, for three different stellar masses which include tracks covering a range of core masses.
For each mass we emphasise the CME model track with a solid black line.
During this time, stars expand as they cross the Hertzsprung-Gap (HG) and then ascend the Red Giant Branch (RGB),
although massive stars ($M\gtrsim14~\Msun$) ignite helium on the HG before reaching the RGB.
The surface properties at high mass ($M\gtrsim 15~\Msun$) should be considered with caution due to the methods used to avoid convergence issues at the Eddington limit (Appendix \ref{append:convergence}).
Our grid also includes stripped stars which are more compact and hotter than non-stripped stars.
As the core mass fraction increases towards 1, the stars move further away from the Hayashi line as the envelope becomes increasingly unable to sustain a giant structure due to its diminishing mass.
These stripped stars are in their pre-helium main sequence phase.

\begin{figure}
 \centering
\includegraphics[width=\linewidth,height=0.9\textheight,keepaspectratio]{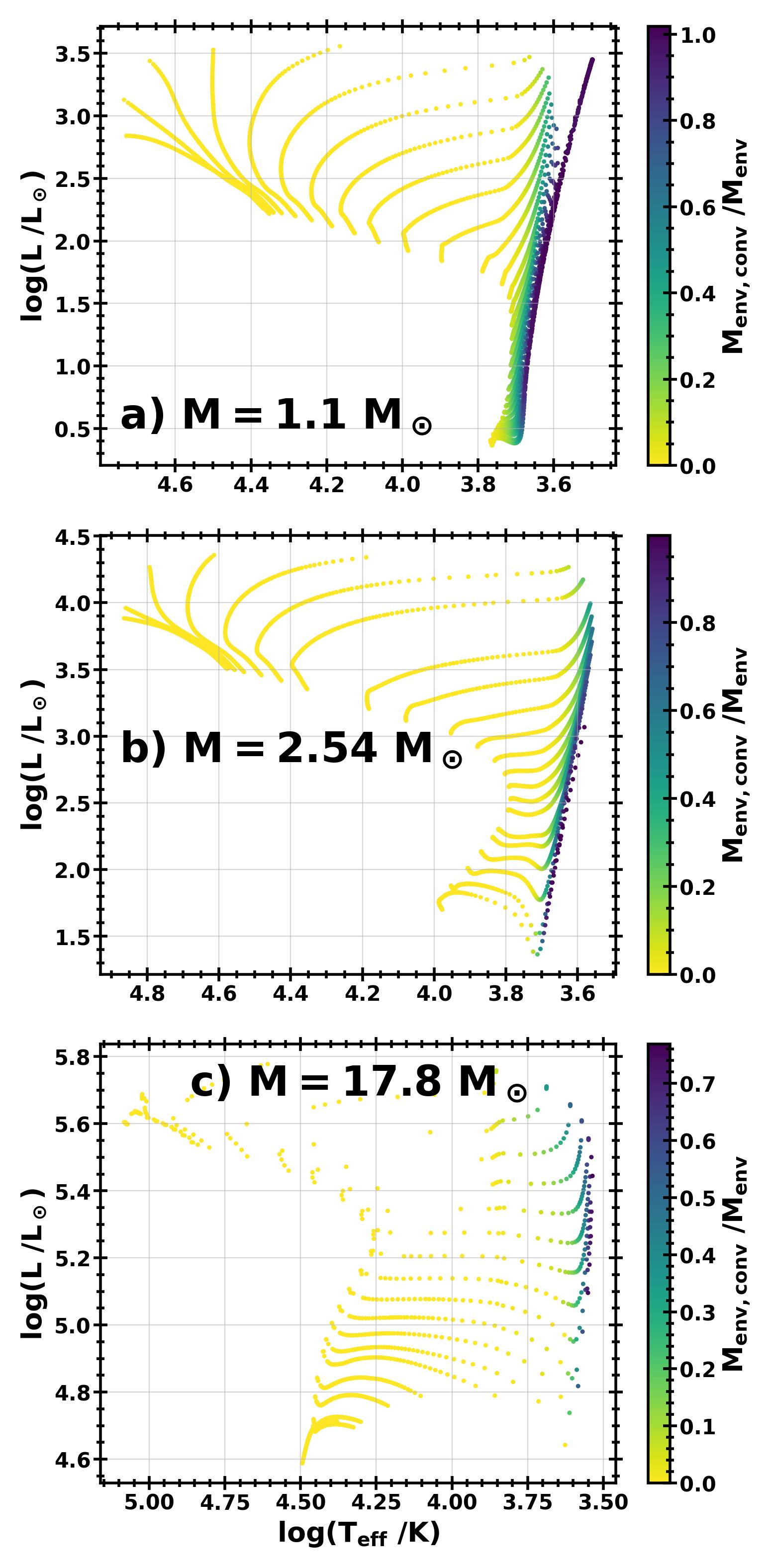}
    \caption{Fig.~\ref{fig:GB_HR} but the tracks are coloured by their convective envelope mass fraction, $M_\mathrm{env,conv}/M_\mathrm{env}$. The growth of the convective envelope indicates the transition from the Hertzsprung Gap (HG) to the Red Giant Branch (RGB).}
    \label{fig:GB_HR_conv_env}
\end{figure}

We produce one table that covers the entire hydrogen-shell burning phase from core hydrogen exhaustion to core helium ignition or WD formation.
However, we conventionally refer to stars by their evolutionary phase split into HG and GB stars.
The transition point from the HG to the GB, known as the base of the giant branch (BGB), is normally identified by looking at the HR diagram.
However, it is useful to precisely define the two phases.
This is particularly important in population synthesis codes which categorise each star in the population by its evolutionary phase.
\citet{hurleyComprehensiveAnalyticFormulae2000} define the BGB to be where the mass fraction of the convective envelope mass exceeds $0.4$ in stars that ignite helium degenerately and where it exceeds $0.33$ in stars that ignite helium gently.
In Fig.~\ref{fig:GB_HR_conv_env} we plot the same HR diagrams as Fig.~\ref{fig:GB_HR} but now coloured by the mass fraction of the envelope which is convective.
In non-stripped stars, there is a clear distinction where the convective envelope rapidly expands in mass as the base of the giant branch.
This leads to the first dredge-up.
Massive stars ($M\gtrsim10~\Msun$) on the HG do have a convective region in their envelope which sits above the core but does not extend to the stellar surface and thus we do not class it as a convective envelope.
\citet{schneiderAgesYoungStar2014} discuss that, in some massive accretors which become blue super-giants, there is a thick convective shell driven by hydrogen-shell burning.
Stripped stars do not reach the giant branch due to their thin, compact envelopes which do not sustain a convective region.
These pre-helium main sequence stars are likely to be rare because stars stripped near the TAMS tend to be massive stars which have a very short hydrogen shell burning phase due to their massive cores which can quickly ignite helium.

\subsection{Helium ignition}
\label{sec:HeIgn}

% \begin{figure} \centering
% \includegraphics[width=\linewidth,height=0.9\textheight,keepaspectratio]{plots/Z0.02_GB_eta.jpeg}
%     \caption{}
%     \label{fig:Mc_eta}
% \end{figure}

\begin{figure*}\centering
\includegraphics[width=\linewidth,height=0.9\textheight,keepaspectratio]{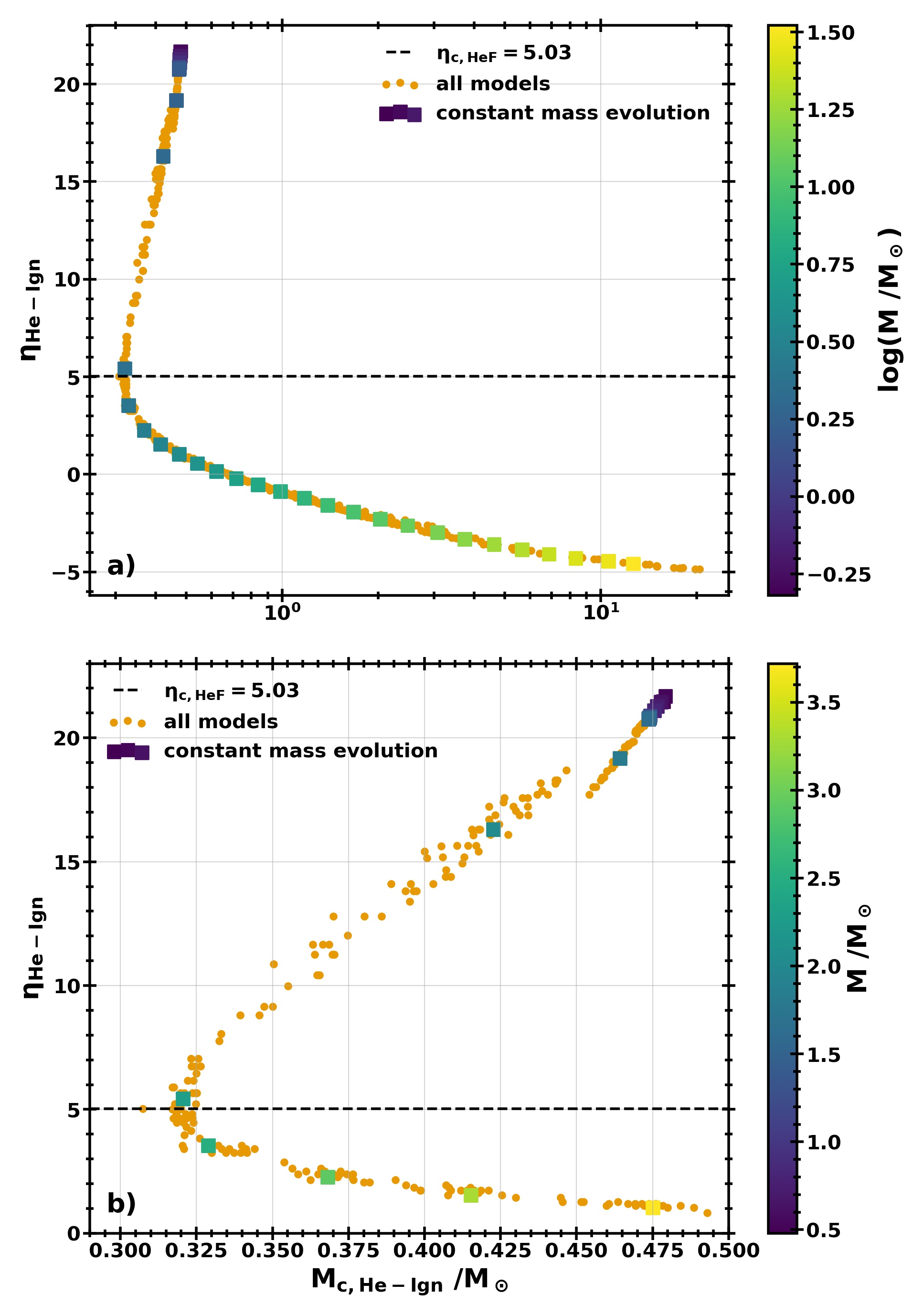}
    \caption{The core mass-central degeneracy relation at helium ignition for (a) all masses and (b) the low mass region where there are degenerate and non-degenerate ignition branches. CME models are plotted with coloured squares corresponding to their stellar mass.}
    \label{fig:He_ign}
\end{figure*}

\begin{figure}\centering
\includegraphics[width=\linewidth,height=0.9\textheight,keepaspectratio]{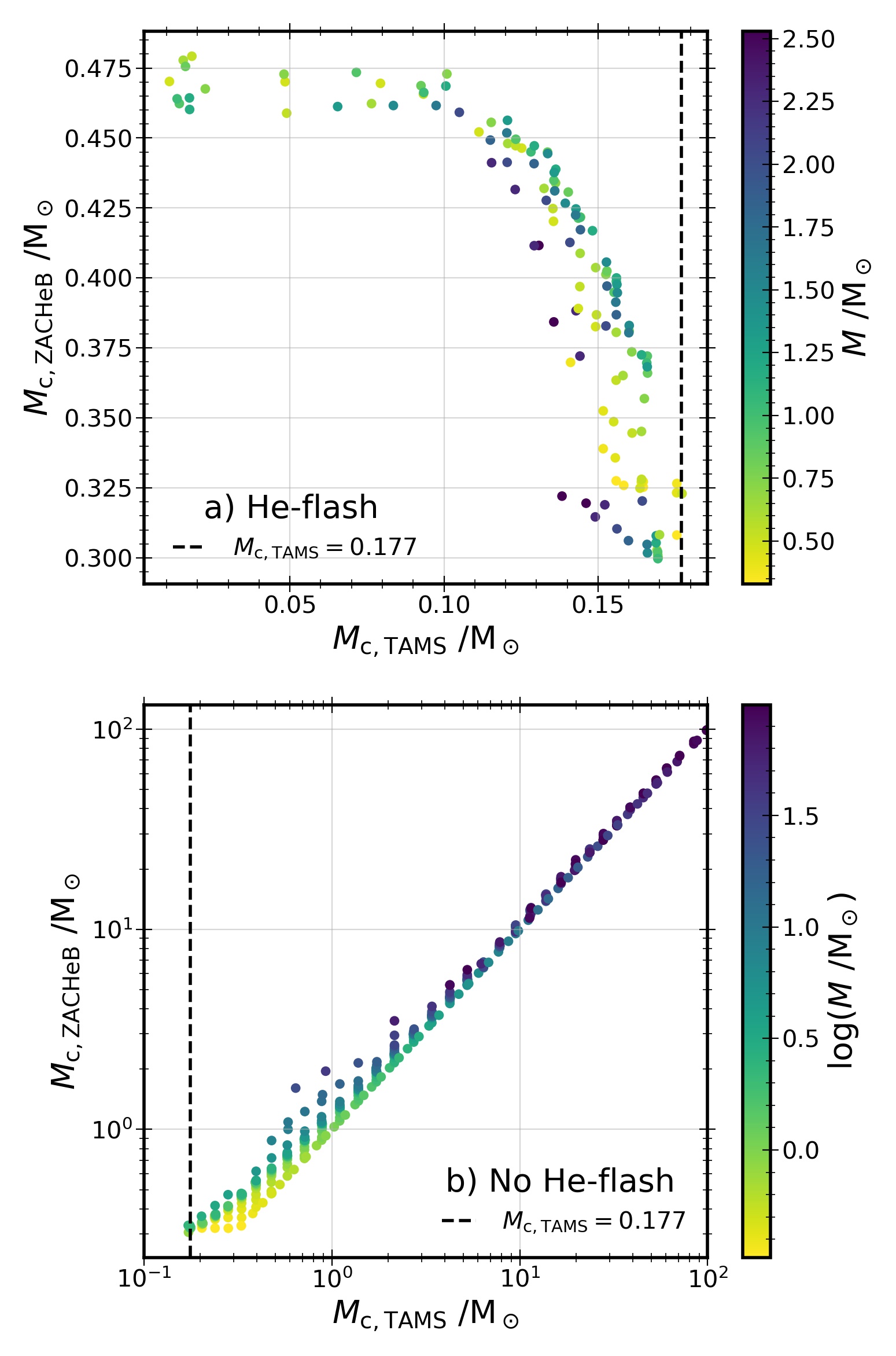}
    \caption{Core mass at the start of helium burning, $M_\mathrm{c,ZACHeB}$, as a function of the terminal age main-sequence core mass, $\McTAMS$, for a) models that He-flash and (b) models that do not He-flash. All models are coloured by the stellar mass. We find that all models with $\McTAMS\gtrsim0.177$ avoid degenerate helium ignition, regardless of envelope mass loss.}
    \label{fig:GB_Mc_growth}
\end{figure}

As stars evolve from core-hydrogen exhaustion to core-helium ignition, the mass and degeneracy of the inert core increases.
Stars during this phase separate into two classes, those with electron-degenerate helium cores, and those with non-degenerate cores.
% Knowing the state of the core is vital for understanding all aspects of the stellar evolution and thus we start by discussing the properties of helium ignition.
We find a tight relation between the core mass and degeneracy at the onset of helium ignition as shown in Fig.~\ref{fig:He_ign}.
% There is a small amount of scatter due to the difficulty in saving the exact point of helium ignition in the model output.
In panel \textit{b} we focus on the low core mass region ($0.3 \lesssim \Mc~/\Msun \lesssim 0.48$) where there is a degenerate ignition branch and a non-degenerate ignition branch.
The bifurcation occurs at a critical central degeneracy $\etacHeF \approx 5$ corresponding to $\Mc \approx 0.32~\Msun$.
In degenerate cores, the higher the degeneracy, the larger the core mass and the more powerful the helium-flash. 

In single star evolution (SSE) there is a minimum initial mass, $M_\mathrm{HeF}\approx 2.3~\Msun$, that is able to ignite helium gently without a flash.
Most stars below this initial mass ignite helium in a strong flash with $\Mc \approx 0.48~\Msun$ and thus attain a similar bolometric luminosity, producing the RGB tip.
However, in a narrow initial mass range, $1.7\lesssim M \lesssim 2.3~\Msun$,  the degeneracy is somewhat reduced leading to a mild He-flash at lower luminosity.
This occurs at all metallicities, although the transition mass $M_\mathrm{HeF}$ is smaller at lower metallicity \citep[e.g.][]{sweigartDevelopmentRedGiant1990, cassisiRedGiantBranch2016}.
The changes that occur in an observed stellar population when stars with $\Mi<M_\mathrm{HeF}$ begin to populate the GB is known as the RGB phase transition \citep{cassisiRedGiantBranch2016}.

In stars with changing mass, such as interacting binaries, the initial mass is not a useful predictor of degenerate ignition.
Instead, we find that the TAMS core mass, $M_\mathrm{c,TAMS}$, is a strong indicator of future helium ignition.
In Fig.~\ref{fig:GB_Mc_growth} we plot the zero-age core-helium burning (ZACHeB) core mass as a function of $\McTAMS$, split into models that He-flash (panel \textit{a}) and those that do not (panel \textit{b}).
We find that all He-flashing models start with $M_\mathrm{c,TAMS}\lesssim 0.165~\Msun$ regardless of the total stellar mass.
Models with more massive cores at the TAMS always avoid the He-flash, regardless of how much of their envelope is lost.
Mass loss impacts the hydrogen-burning shell and thus reduces the rate of core mass growth, but never by enough to cause the core to become degenerate before helium ignition.
Similarly, almost all models with $M_\mathrm{c,TAMS}\lesssim 0.165~\Msun$ undergo the He-flash even when mass is accreted.
Mass accretion increases the rate of the core mass growth but not enough to prevent degenerate ignition.
There is an exception in the case of the strong mass accretion discussed in section \ref{sec:InterpVariables}.
Models with $M>2.5~\Msun$ that accrete such that $\Mc/M\lesssim0.05$, undergo a readjustment phase during which the core degeneracy is decreased and the core mass rapidly grows.
Via this mechanism, cores which are otherwise destined to He-flash, rapidly adjust and consequently ignite helium gently.
We thus find that stars with $M\gtrsim 3~\Msun$ will never He-flash, regardless of the initial stellar mass.

In stars that ignite helium gently (Fig.~\ref{fig:GB_Mc_growth} b), the envelope mass has an impact on the core mass at helium ignition when $\McTAMS \lesssim 1~\Msun$.
At larger total stellar mass, the core grows more due to a more efficient hydrogen-burning shell.
The yellow points correspond to stripped stars which cannot grow their cores due to a lack of hydrogen rich envelope available to the burning shell.
At higher core masses, the hydrogen shell burning phase is so short that there is little core mass growth and $\McTAMS\approx M_\mathrm{c,ZACHeB}$ at all stellar masses.

\subsection{Luminosity-core mass relation}

\begin{figure*}
 \centering
\includegraphics[width=\linewidth,height=0.9\textheight,keepaspectratio]{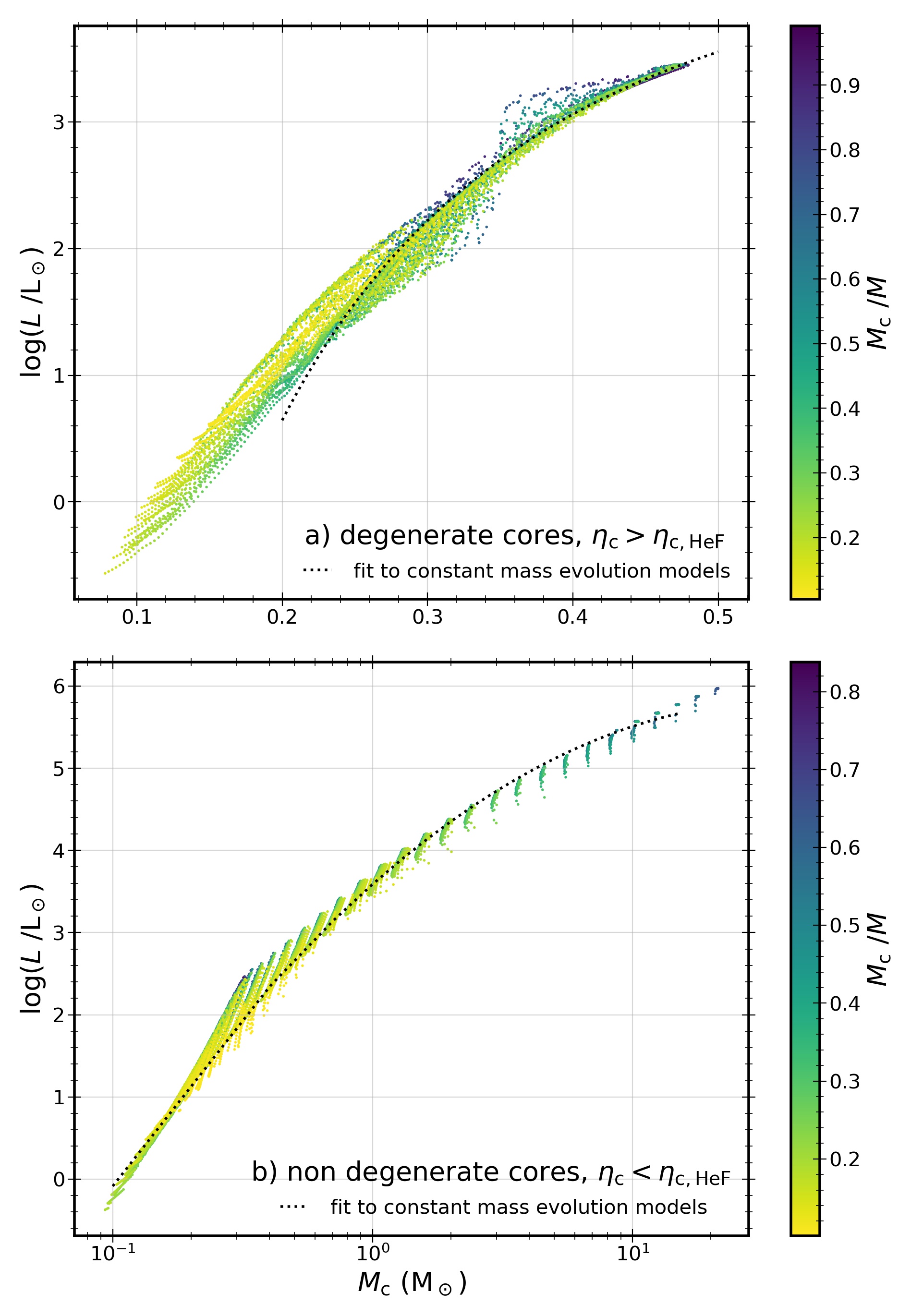}
    \caption{Stellar luminosity on the first giant branch, $L$, as a function of core mass, $\Mc$, in stars with a) degenerate cores and b) non-degenerate cores. The core mass fraction, $\Mc/M$, is indicated by the colour of the scatter points.
    Also plotted in dotted, black lines are fits to the constant-mass evolution models (Eqs.~\ref{eq:L(Mc)_degen} and \ref{eq:L(Mc)_non_degen}).}
    \label{fig:lum_core_mass_relation}
\end{figure*}

In SSE, it is known that the GB stellar luminosity is primarily set by the core mass, $\Mc$, which decides the efficiency of the hydrogen-burning shell \citep[e.g.,][Fig.~11]{hurleyComprehensiveAnalyticFormulae2000}.
Fig.~\ref{fig:lum_core_mass_relation} shows the luminosity vs the core mass of all GB models in the grid, i.e. those with $M_\mathrm{env,conv}/M_\mathrm{env}>0.4$.
The core mass-luminosity relation is split into two branches.
There is a degenerate core branch (\textit{a}), which forms the RGB tip at a maximum luminosity $L_\mathrm{RGB tip}\sim2,500~\Lsun$, and a non-degenerate core branch (\textit{b}).
Giant stars with degenerate helium cores and a hydrogen burning shell source are thought to be well fit by a single core mass- luminosity relation \citep{refsdalShellSourceBurning1970}. 
We thus fit the tip of the RGB with a second order polynomial to obtain,
\begin{equation}\label{eq:L(Mc)_degen}
    \log(L~/\Lsun) = -7.4\log(\Mc~/\Msun)^2-0.1\log(\Mc~/\Msun)+4.2.
\end{equation}
Note that whilst all GB models are plotted in Fig.~\ref{fig:lum_core_mass_relation}, only the CME models were used to obtain the fit.
Deviations from this fit are at low core masses, $\Mc\lesssim0.25~\Msun$.
We also find a population of stripped models with core masses $0.35\lesssim\Mc~/\Msun\lesssim0.45$ that have increased luminosities with respect to the relation.
We thus find that whilst a single core mass- luminosity relation is a decent first approximation, there is a secondary impact of the envelope mass on the stellar luminosity.

We also give an approximate fit to our non-degenerate GB CME models,
\begin{equation}\label{eq:L(Mc)_non_degen}
    \log(L~/\Lsun) = -0.9\log(\Mc~/\Msun)^2+2.8\log(\Mc~/\Msun)+3.6.
\end{equation}
Again only the CME models were used to obtain the fit but all models are plotted in Fig.~\ref{fig:lum_core_mass_relation} \textit{b}.
This can be used to obtain an approximate GB luminosity at a given core mass however for each star the luminosity increases as it climbs the GB.
We thus find that in both stars with degenerate and stars with non-degenerate cores the luminosity is a function of the core mass and the total mass, $L=L(\Mc, M)$.

\section{Discussion}
\label{sec:Discussion}

We now turn our attention to how our model grid can be used to produce an interpolation table for use in population synthesis codes such as \textit{binary\_c}.
The \textit{binary\_c} algorithm evolves stars by taking timesteps. 
Within the \textit{MINT} library, the interpolation variables define the state of the star in a vector.
For example, on the MS the star is defined by $\textbf{x} = (M, X_\mathrm{c})$, where $M$ is the stellar mass and $X_\mathrm{c}$ is the central hydrogen abundance.
In the next evolutionary phase (HG\&GB), the state is defined by a three dimensional vector, $\textbf{x} = (M, \Mc/M, \etac)$.
The state is evolved by integrating the derivatives of the interpolation variables with respect to time.
Given the state of a star at some time, $\textbf{x}(t)$, we can compute the state some time $dt$ later,
\begin{equation}
\label{eq:star_integration}
    \textbf{x}(t+dt) = \textbf{x}(t) + \frac{d\textbf{x}}{dt}dt,
\end{equation}
where the timestep, $dt$ is sufficiently small.
Thus, we provide first derivatives of $\Mc/M$ and $\etac$ in the table.
The rate of change of the stellar mass, $M$, is defined by wind mass loss and binary mass-transfer.
The stellar properties are found by interpolating at each timestep.
Details of all the data available in the \textit{MINT} tables are included in Appendix \ref{append:MINTTable}.
We use a number of post-processing steps to prepare the table for use in a population-synthesis code.

% \subsection{Numerical Smoothing}

% The evolution of the time-proxy, $\etac$, with time is non-linear and the first derivative, $\frac{d\etac}{dt}$, varies by orders of magnitude in a particular evolutionary track.
% Such changes in the gradient leads to inaccuracies when integrating the stellar age. 
% To prevent such errors in the lifetimes we use mean smoothing on the time-proxy save target times used to calculate the derivative. 
% The smoothing is applied locally, only including 5 points either side of each target however it is applied three times to smooth out gradient changes.

% The core mass must also be integrated to evolve the stellar state.
% We can see from Fig.~\ref{fig:GB_tracks} that there are cases when the core mass is approximately constant.
% During times of low growth, the core mass time evolution exhibits step-like behaviour due to the discretization of mass cells in MESA models. 
% Step-like behaviour leads to extreme gradient changes in the derivative of core mass with respect
% to time and thus also lead to large errors during integration. 
% To increase the accuracy of the core mass integration we also use mean smoothing on the core mass values.

\subsection{Interpolation variable remapping}
\label{sec:InterpVariableRemapping}

\begin{figure*}
 \centering
\includegraphics[width=\linewidth,height=0.9\textheight,keepaspectratio]{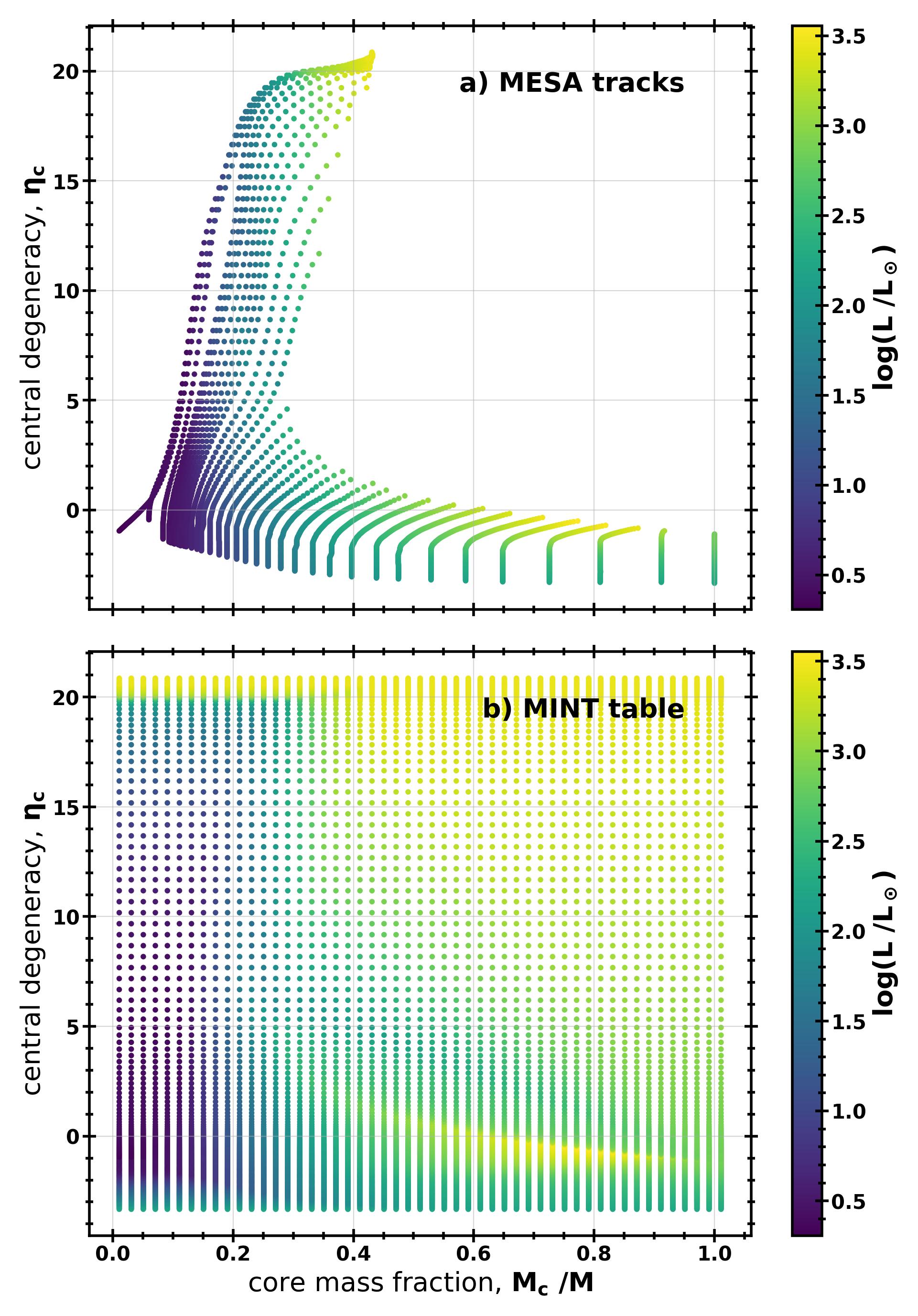}
    \caption{An example of the interpolation variable remapping process for $M=1.1~\Msun$ where the quantity being interpolated is the logarithm of the stellar luminosity, $\log(L~/\Lsun)$. The \textit{MESA} tracks (\textit{a}) are mapped onto an orthogonal grid in the core mass fraction- central degeneracy parameter space (\textit{b}).}
    \label{fig:GB_remapping}
\end{figure*}

\begin{figure}
 \centering
\includegraphics[width=\linewidth,height=0.9\textheight,keepaspectratio]{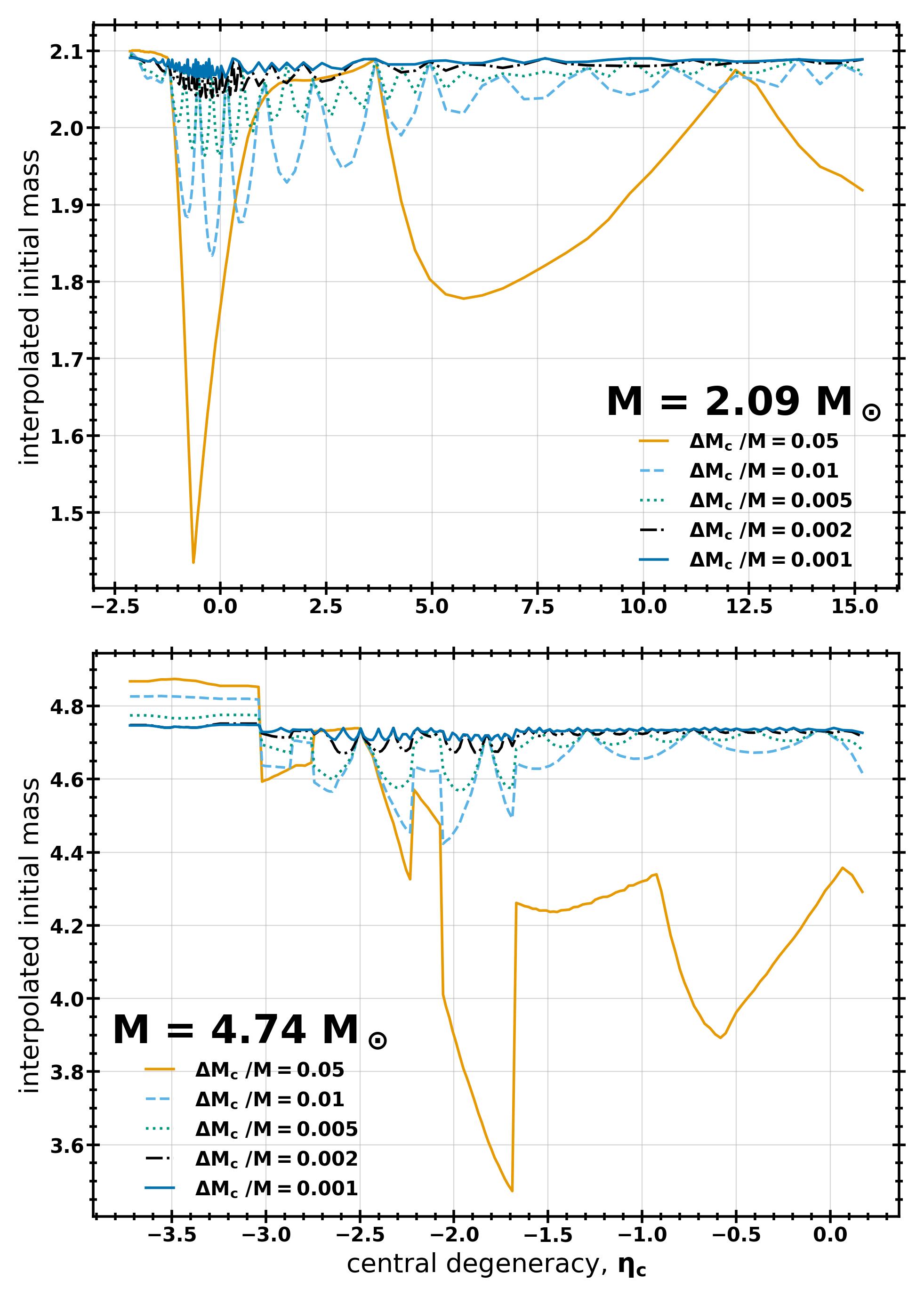}
    \caption{The interpolated initial mass resulting from the interpolation variable remapping process onto an orthogonal grid for \textit{MESA} tracks with a) $M=\Mi = 2.09~\Msun$ and b) $M=\Mi=4.74~\Msun$.
    The impact of numerical diffusion decreases as the core mass fraction resolution of the orthogonal grid is increased.}
    \label{fig:remapping_numerical_diffusion}
\end{figure}

The \textit{MINT} library utilises linear interpolation for the sake of speed and scalability.
The \textit{librinterpolate} library (\url{https://gitlab.com/rob.izzard/librinterpolate}) is used to interpolate on N-dimensional, orthogonal, complete data sets.
The MS grid naturally produces an orthogonal table because all masses have a data point for each target value of the central hydrogen abundance.
Beyond the MS, the stellar evolution tracks produce non-orthogonal interpolation tables (Fig.~\ref{fig:GB_tracks}). 
The interpolation library has been updated to handle non-orthogonal tables by remapping the data to a fully orthogonal table.
However, it is desirable to remap the data prior to being loaded by \textit{MINT} so that fewer calculations have to be computed on-the-fly.
This reduces the load and preparation time of tables in \textit{binary\_c} and also allows for easier inspection of the remapping process to check for unwanted interpolation artifacts.
We use the \textit{scipy.interpolate.griddata} Python function to interpolate our unstructured data onto an orthogonal grid.
We use linear interpolation within the bounds of the original data and then fill the outside area using nearest neighbour extrapolation.
In Fig.~\ref{fig:GB_remapping} we show an example of the remapping process for models with $M=1.1~\Msun$ in which the quantity being remapped is the convective envelope mass fraction.

However, the remapping process leads to a loss of accuracy via numerical diffusion.
We can minimise the numerical diffusion by increasing the resolution of the core mass fraction values to which the data are remapped.
In our tables we include the column INITIAL\_MASS which identifies the MESA track from which data originate.
In the remapped table, the value of INITIAL\_MASS should be constant along the path of the original \textit{MESA} tracks in the parameter space of the interpolation coordinates.
A good way to test the extent of numerical diffusion is to compare the value of INITIAL\_MASS when interpolating from the remapped table along these tracks.
In Fig.~\ref{fig:remapping_numerical_diffusion} we compare the interpolated values of INITIAL\_MASS along two CME tracks on the giant branch with \textit{a}) $M=2.09~\Msun$ and \textit{b}) $M=4.74~\Msun$.
Were there no numerical diffusion, the interpolated initial mass should be constant and equal to the initial mass, $M_\mathrm{i}$.
Increasing the resolution of the remapped core-mass fraction decreases the numerical diffusion, but also increases the size of the remapped table and thus increases storage.
The chosen remapped core-mass fraction resolution is therefore a compromise between the impact of numerical diffusion and the size of the \textit{MINT} table.
We choose a value of $\Delta\Mc/M=0.002$.

\begin{figure}
 \centering
\includegraphics[width=\linewidth,height=0.9\textheight,keepaspectratio]{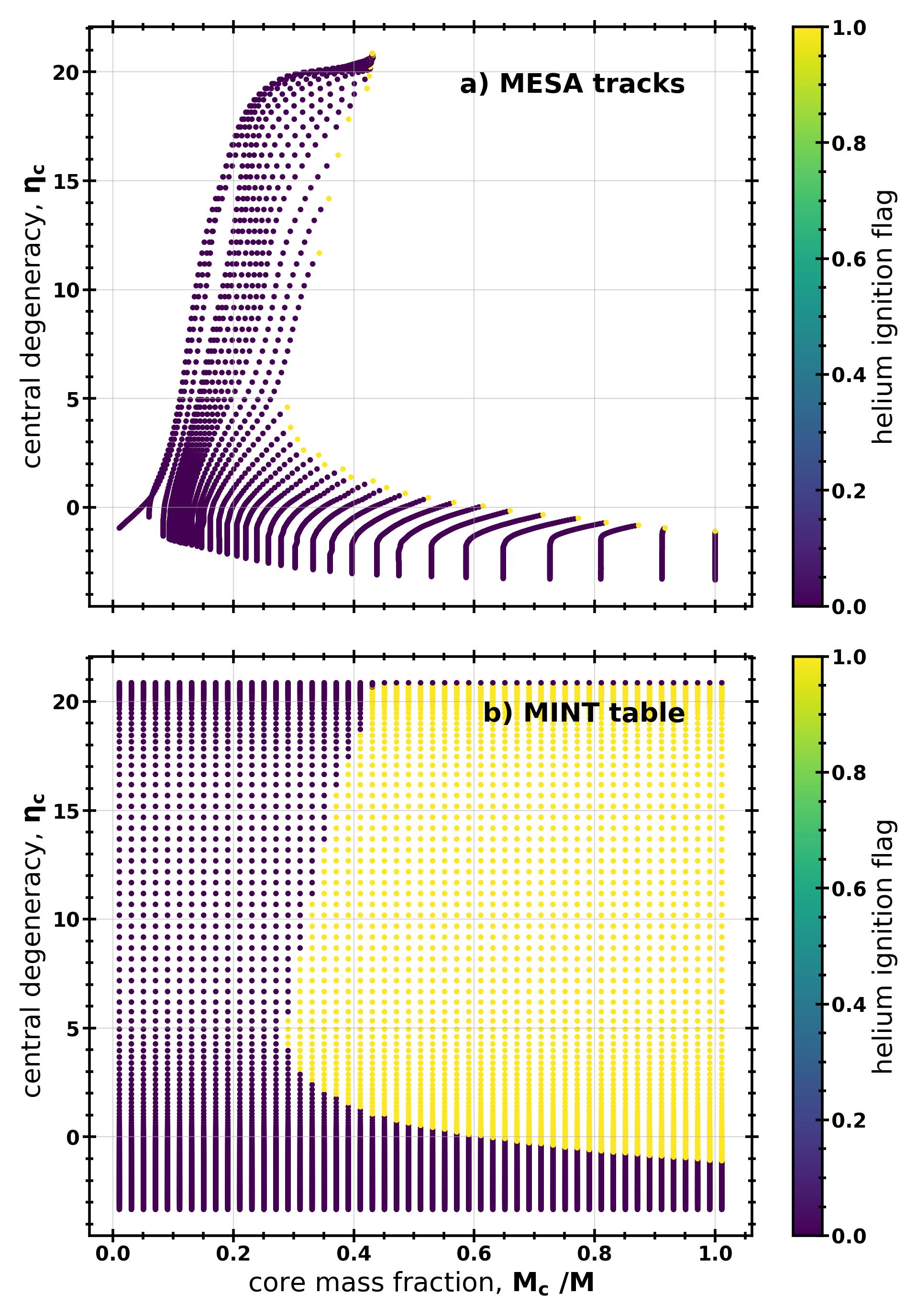}
    \caption{The helium ignition flag for $M=1.1~\Msun$ in \textit{a}) the MESA tracks and \textit{b}) the MINT table. The parameter space section coloured in yellow is where helium has ignited and the section coloured in dark purple is where helium is yet to ignite. We use the post-processing method described in section \ref{sec:InterpVariableRemapping} to restore the curved ignition boundary after interpolation coordinate remapping onto an orthogonal grid.}
    \label{fig:GB_helium_ignition_remapping}
\end{figure}

To evolve a star through all its evolutionary phases, the \textit{MINT} library must instruct \textit{binary\_c} to switch to the next table at the end of an evolutionary phase.
The end of the MS has a simple transition point at core hydrogen exhaustion, for example $\Xc\lesssim10^{-6}$.
Finding the onset of core-helium burning is not so simple because the central degeneracy at helium ignition varies depending on the core mass (Fig.~\ref{fig:He_ign}).
Thus, instead of a single value of the central degeneracy time-proxy at which the transition occurs there is a curve in the core mass- central degeneracy parameter space (Fig.~\ref{fig:He_ign}).
We use a flag to denote the ignition of helium which is set to 1 at the end of each track and 0 elsewhere.
However, due to this being a binary quantity, this boundary is not well maintained during the interpolation variable remapping onto an orthogonal grid.
Thus, we add a post-processing step after the interpolation coordinate remapping to restore the helium ignition boundary using the curve in Fig.~\ref{fig:He_ign}.
We use the points on the curve to produce an interpolation function for the core mass at helium ignition, $M_\mathrm{c,ign}$, as a function of the central degeneracy at helium ignition, $\eta_\mathrm{c,ign}$,
\begin{equation}
    M_\mathrm{c,ign} = f(\eta_\mathrm{c,ign}).
\end{equation}
Then, at each point in the orthogonal grid, we set the helium ignition flag to 1 if $\Mc>=f(\etac)$ and 0 if $\Mc<f(\etac)$.
We show the results of this post-processing step in Fig.~\ref{fig:GB_helium_ignition_remapping} for total mass $M=1.03~\Msun$.
The final \textit{MINT} table is produced after these post-processing steps.

\subsection{Testing MINT}\label{sec:MINTTesting}

\begin{figure*}
 \centering
\includegraphics[width=\linewidth,height=0.9\textheight,keepaspectratio]{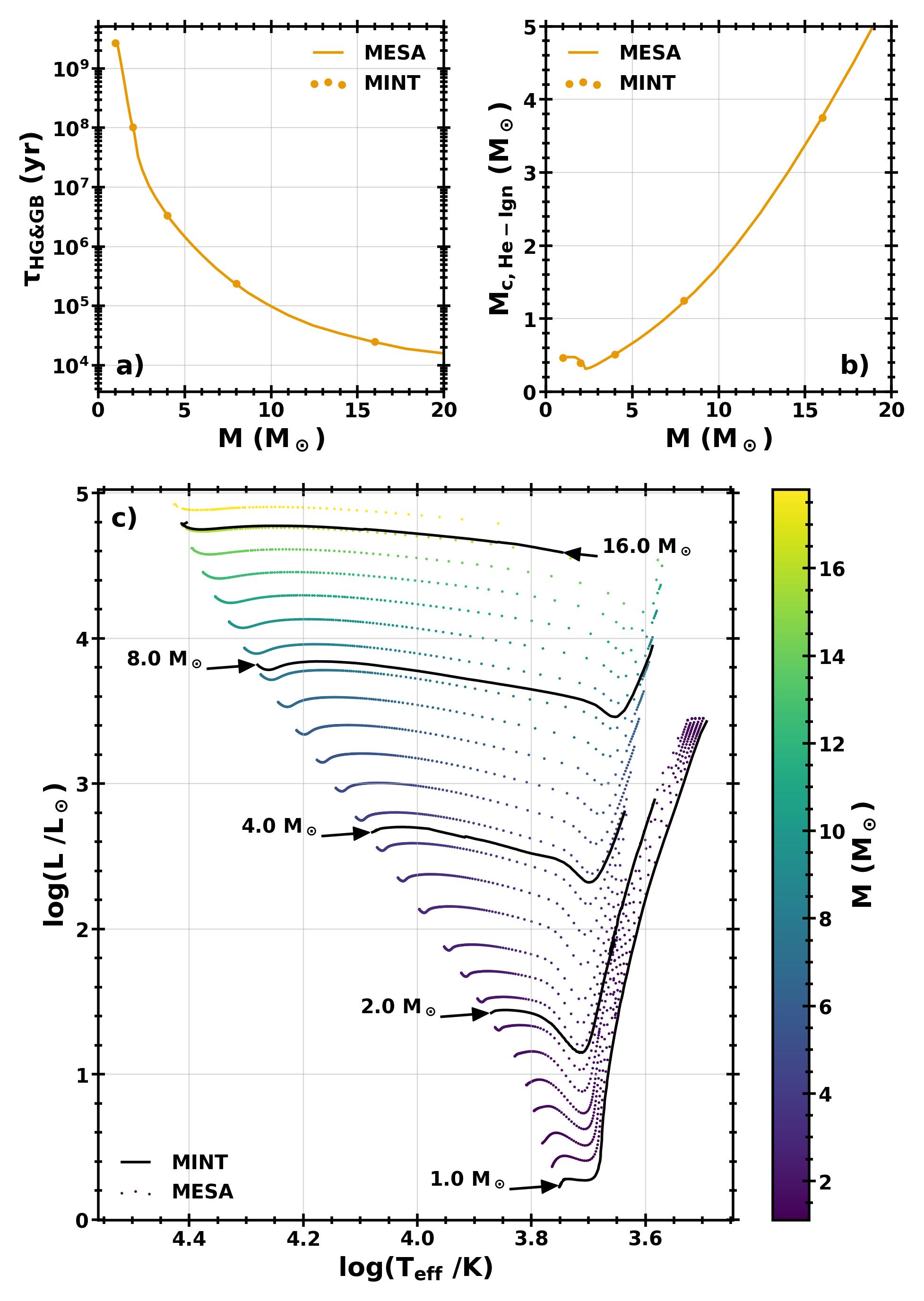}
    \caption{A comparison of integrated \textit{MINT} tracks to \textit{MESA} tracks at constant mass and $Z=0.02$. \textit{a)}) the HG and GB lifetime, $\tau_\mathrm{HG\&GB}$, as a function of stellar mass. \textit{b}) the core mass at helium ignition, $M_\mathrm{c,He-Ign}$, as a function of stellar mass. \textit{c}) The HR diagram for all tracks. The \textit{MINT} tracks are plotted in solid black and are labelled by their stellar masses. The \textit{MESA} tracks are plotted in scatter points coloured by their stellar masses.}
    \label{fig:MINT_HR}
\end{figure*}

\begin{figure*}
 \centering
\includegraphics[width=\linewidth,height=0.9\textheight,keepaspectratio]{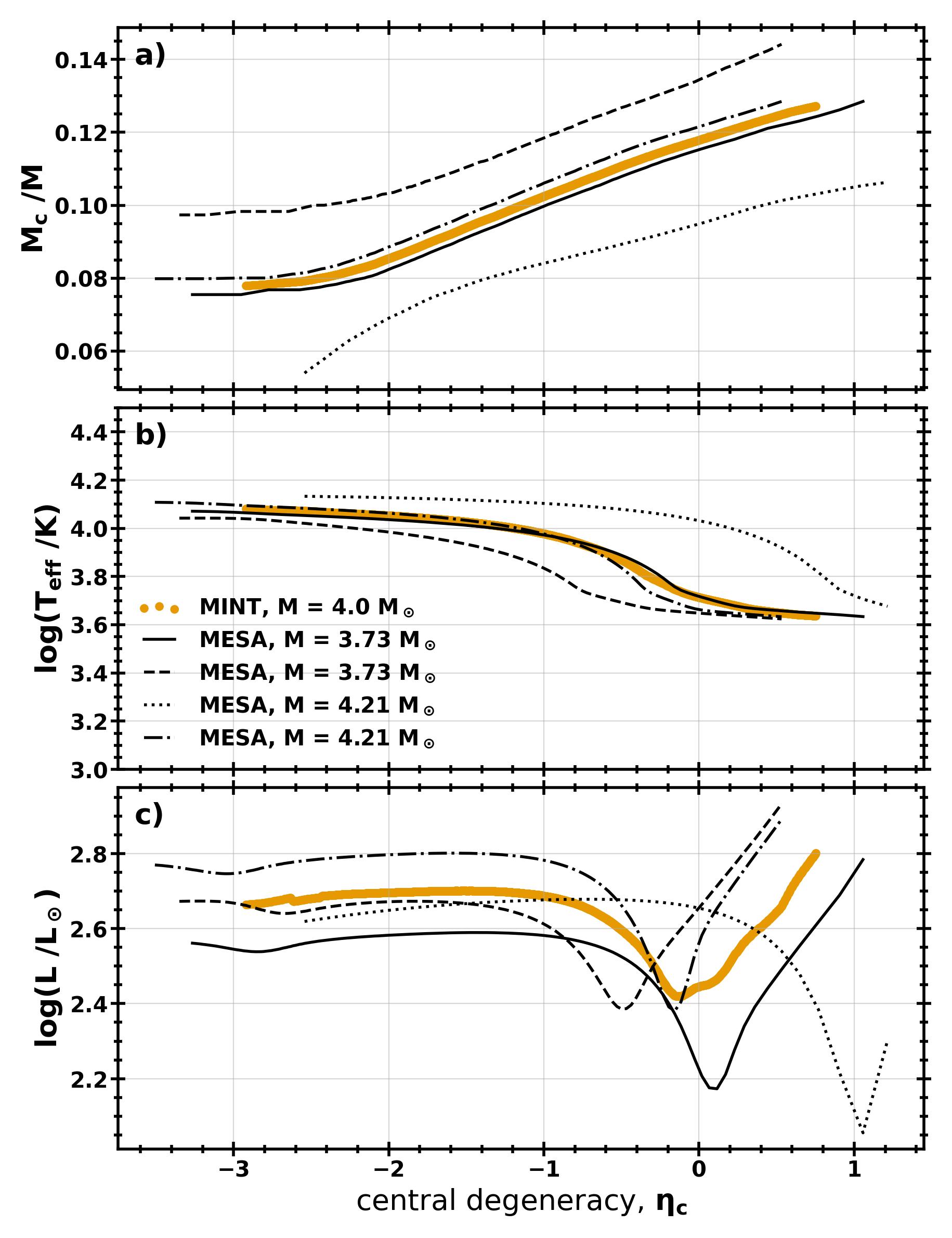}
    \caption{Comparison of a $4~\Msun$ \textit{MINT} track (thick, orange) to our four \textit{MESA} tracks (black) that are closest in core mass fraction and stellar mass. \textit{a}) The core mass fraction, $\Mc$/M. \textit{b}) The effective temperature, $\Teff$. \textit{c}) The stellar luminosity, $L$. Interpolation as a function of the central degeneracy, $\etac$, works for the effective temperature but smears out the luminosity minimum at the base of the giant branch due to the highly non-linear behaviour.}
    \label{fig:MINT_INT_EX}
\end{figure*}

\begin{figure}
 \centering
\includegraphics[width=\linewidth,height=0.9\textheight,keepaspectratio]{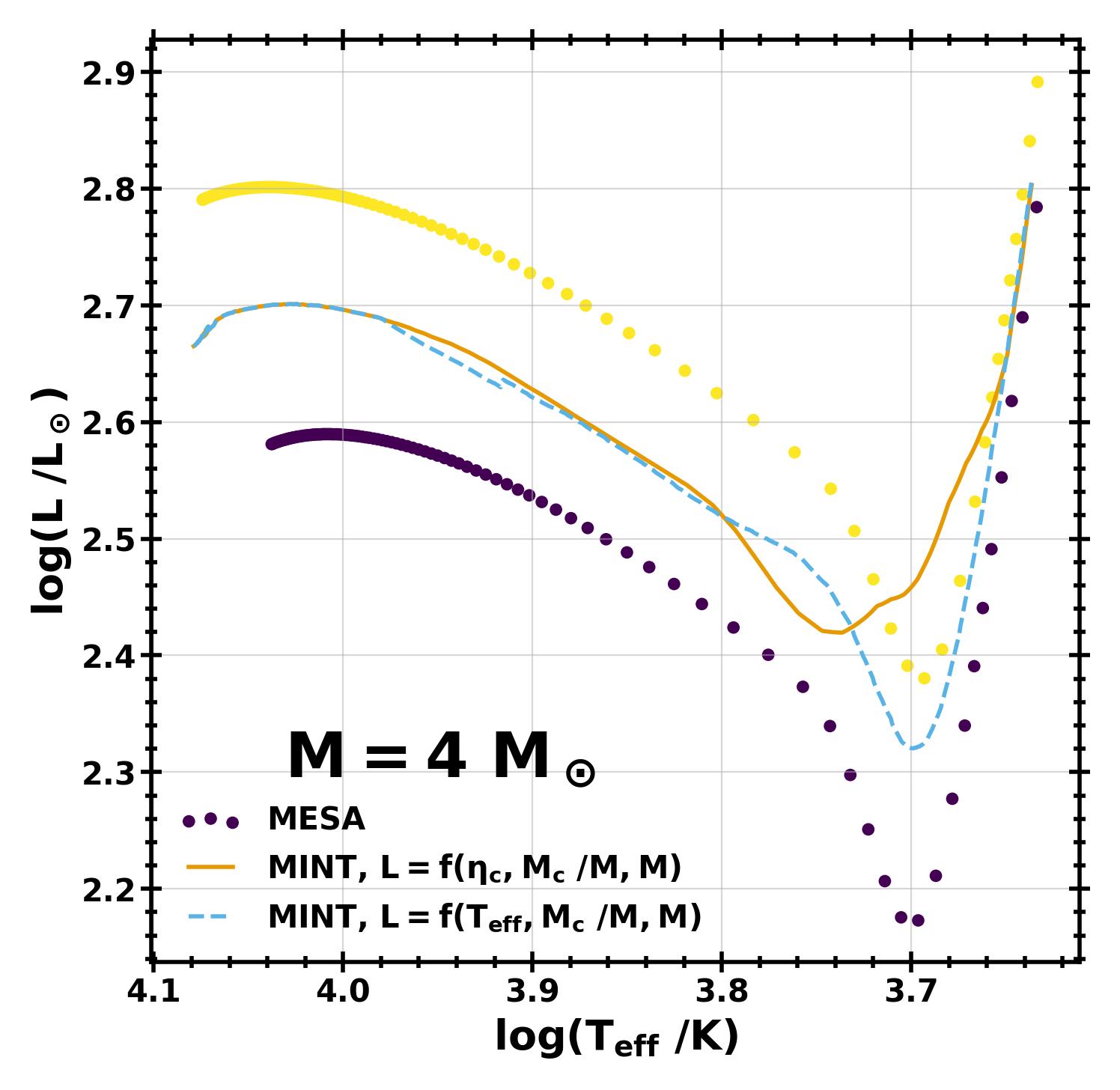}
    \caption{A comparison of the \textit{MINT} stellar luminosity for a $4~\Msun$ star, integrated as a function of the central degeneracy, $\etac$, (solid), and as a function of the effective temperature, $\Teff$, (dashed). We also show our \textit{MESA} tracks for the closest stellar masses, $3.73$ and $4.21~\Msun$.}
    \label{fig:MINT_LUM_INT}
\end{figure}

% We undergo some testing to ensure that we can reconstruct the original \textit{MESA} evolutionary tracks.
% This ensures that there is no significant loss of information between the \textit{MESA} output files and the \textit{MINT} interpolation table.
To mimic the \textit{binary\_c} algorithm, we have written an integration code in Python which uses the forward Euler method to evolve the stellar state vector (Eq.~\ref{eq:star_integration}).
The \texttt{scipy} \texttt{interpolate} \texttt{RegularGridInterpolator} method is used to interpolate quantities from the \textit{MINT} table.
To test the \textit{MINT} method we produce integrated evolution models for initial masses, $M = $ 1, 2, 4, 8 and 16$~\Msun$ without mass-loss.
In Fig.~\ref{fig:MINT_HR} we show that the \textit{a}) HG \& GB lifetime, $\tau_\mathrm{HG\&GB}$, and \textit{b}) core mass at helium ignition, $M_\mathrm{c,He-Ign}$, are well reproduced for all masses by comparing to the original \textit{MESA} models.
We also take a closer look at how the effective temperature, $\Teff$, and stellar luminosity, $L$ are reproduced using this method. 

In Fig.~\ref{fig:MINT_INT_EX} we compare the integrated track for $4~\Msun$ to the four \textit{MESA} tracks which are closest in the 
mass and core-mass fraction parameter space.
The effective temperature (\textit{b}) is monotonic with the time-proxy, $\etac$, and interpolation in the state vector is able to reproduce the expected evolutionary behaviour throughout this phase.
However, the luminosity (\textit{c}) is highly non-linear with respect the time-proxy, $\etac$.
We see that the luminosity minimum at the base of the RGB occurs at a different $\etac$ in each track which leads to a smeared out minimum in the integrated track.
Thus, the central degeneracy is not a good time-proxy for interpolating the stellar luminosity due to the unaligned luminosity profiles.
To fix the luminosity profile, we perform a secondary step to interpolate the luminosity as a function of the effective temperature, core mass fraction and stellar mass.
This works in this phase because the effective temperature is monotonic and the base of the RGB occurs at approximately the same effective temperature at all masses.
Note that because the effective temperature was interpolated as a function of the state vector, this is still interpolating as a function of the state vector but with an intermediate transformation step,
\begin{equation}\label{eq:lum_int}
    L = f'(\Teff,\Mc/M,M) = f(\etac,\Mc/M,M) =  f(\textbf{x}).
\end{equation}
We use the same methods to remap the luminosity on to an orthogonal grid and interpolate as a function of $\Teff$, $\Mc/M$ and $M$.
In Fig.~\ref{fig:MINT_LUM_INT} we compare the result of the additional interpolation step with the first interpolation and the original \textit{MESA} tracks for masses $M=3.73$ and $4.21~\Msun$.
With the second interpolation as a function of $\Teff$ we are able to reconstruct the HR diagram for the GB.
In Fig.~\ref{fig:MINT_HR} \textit{c}) we show the HR diagram for all the integrated \textit{MINT} tracks compared the original \textit{MESA}.
We thus see that our method is able to reproduce the HR diagram.

\section{Limitations}
\label{sec:Limitations}

We now highlight what we consider to be the most important limitations of our method.

\subsection{Main sequence rejuvenation}
\label{sec:MSRejuvenation}

The existence and size of a convective core during core-hydrogen burning strongly depends on the stellar mass which compresses and heats the stellar center.
This gives the mass-luminosity relation for MS stars, $L\propto M^x$, where $x\approx 3.5$ \citep{kuiperEmpiricalMassLuminosityRelation1938}.
Thus, if mass is accreted onto a MS star, its thermal-equilibrium luminosity increases and the convective region expands, or is created if the core is originally radiative.
Rejuvenation occurs because additional unprocessed material is mixed into the core, causing the hydrogen mass fraction to increase and the star to appear younger \citep{neoEffectRapidMass1977, hellingsPhenomenologicalStudyMassive1983, hellingsPostRLOFStructureSecondary1984, schneiderAgesYoungStar2014, schneiderEvolutionMassFunctions2015, schneiderRejuvenationStellarMergers2016}.
This leads to populations of blue stragglers \citep{rasioMinimumMassRatio1995, sillsEvolutionStellarCollision1997, sillsEvolutionStellarCollision2001, vanbeverRejuvenationStarburstRegions1998, mapelliRadialDistributionBlue2006, glebbeekEvolutionStellarCollision2008, ferraroDynamicalAgeDifferences2012}.
\citet{hellingsPhenomenologicalStudyMassive1983,hellingsPostRLOFStructureSecondary1984} found that after rejuvenation, the internal chemical structure of the star is almost identical to that of a single star with the new mass and central hydrogen abundance.
However, they use the Schwarzschild criterion which does not consider the impact of mean molecular weight barriers.
As hydrogen is burnt in the center, a composition gradient gradually forms, dividing the star into a core and envelope.
If the core is sufficiently separated from the envelope by a mean molecular weight barrier, rejuvenation will not occur.
The existence of some composition gradient may also lead to partial rejuvenation where the convective core mass increases but not to the size one would expect for a single star of the same mass.
\citet{braunEffectsAccretionMassive1995} found that full rejuvenation of massive accretors can occur via the formation of a semi-convection region provided the semi-convection efficiency parameter is large enough.
However, at smaller semi-convective efficiencies only partial rejuvenation occurs.
In addition, the critical efficiency parameter is a function of the core hydrogen abundance, the mass of matter accreted and the initial mass of the accreting star.
\citet{schneiderPresupernovaEvolutionFinal2024} computed models with a stronger semi-convection efficiency than \citet{braunEffectsAccretionMassive1995} and found that accretors with $\Xc> 0.05$ (almost) fully rejuvenate when more than 10\% of the initial stellar mass is accreted.
The point along the MS at which full rejuvenation no longer occurs is currently unknown and is likely dependent on stellar mass and metallicity, as well as convective boundary mixing.

Partial rejuvenation directly impacts the composition profile outside the core that forms at the TAMS \citep[e.g.][Fig.~5]{braunEffectsAccretionMassive1995}.
This composition profile persists until the hydrogen-burning shell grows past it, unless FDU is deep enough to convert the profile into a step-like function.
By using cores from CME models, our models are applicable for the case of mass transfer after the TAMS or the case of full rejuvenation.
If partial rejuvenation leads to different composition profiles, the ages and core masses of models will be inaccurate, due to the hydrogen-burning shell moving through the composition gradient at a different rate.
However, it is worth noting that our models with convective-envelope overshooting experience deep FDU, which removes any complicated composition profiles.
Although rejuvenation may also have other consequences.
For example, 
\citet{renzoRejuvenatedAccretorsHave2023} found that MS rejuvenation modifies the core-envelope boundary enough to significantly decrease the envelope binding energy which has consequences for common envelope events.

\subsection{Dynamical mass changes}

In this work we produce models that are in thermal equilibrium, with the exception of the deviation away from thermal equilibrium that occurs as stars cross the Hertzsprung gap (section.~\ref{sec:HGThermalEquilibrium}).
Mass changes that occur quicker than the thermal timescale force the star out of thermal equilibrium \citep[][and references therein]{woodsCanWeTrust2011}.
Thus additional prescriptions will be needed to determine the surface properties of these stars until thermal equilibrium is restored.
For example, \cite{lauExpansionAccretingMainsequence2024} investigated the expansion of main sequence stars with initial masses $2-20~\Msun$ at a range of accretion rates and present a prescription for implementing their results in a population synthesis code.

\subsection{Loss of thermal equilibrium on the Hertzsprung-Gap}
\label{sec:HGThermalEquilibrium}

\begin{figure*}
 \centering
\includegraphics[width=\linewidth,height=0.8\textheight,keepaspectratio]{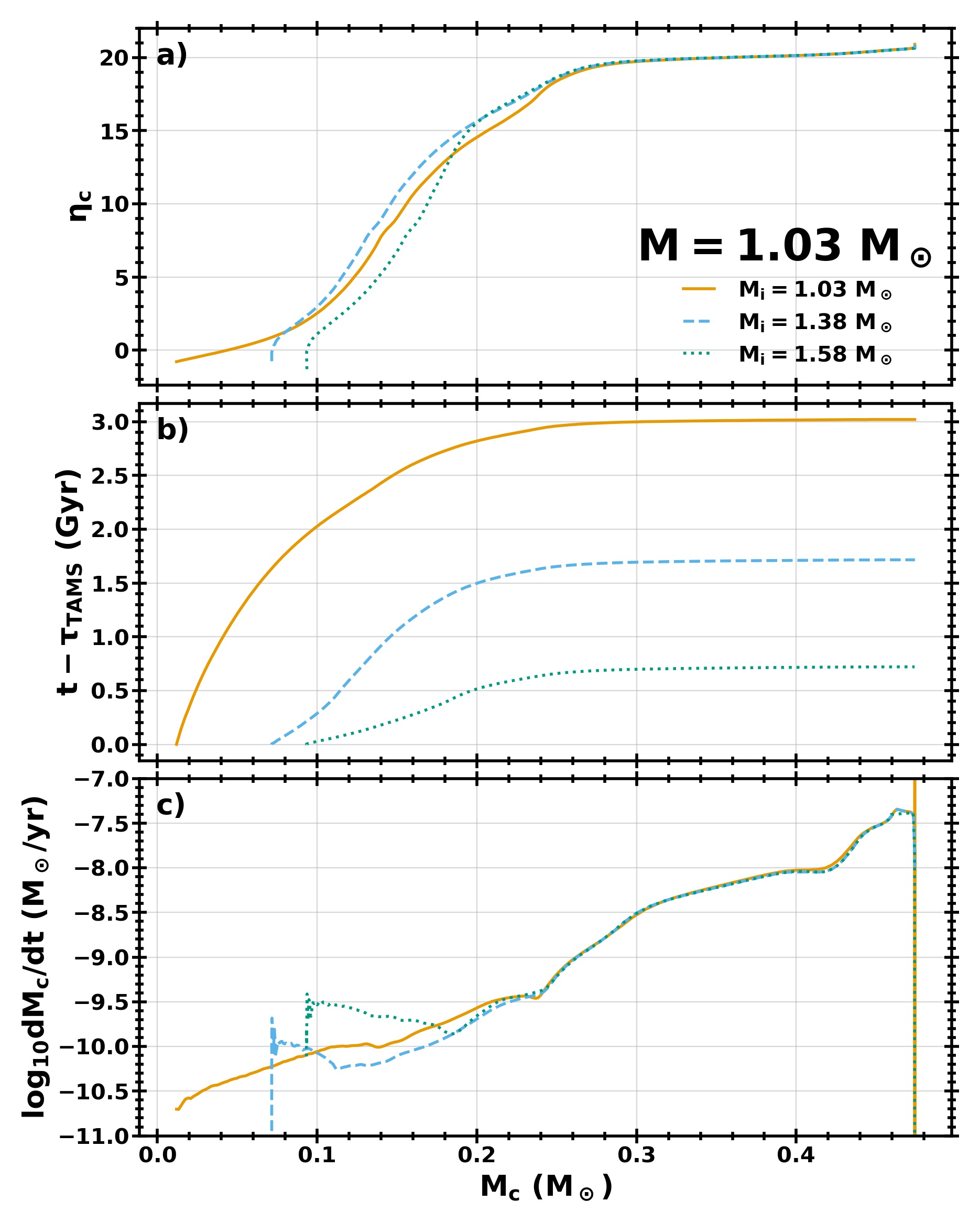}
    \caption{The HG and GB evolution for three models at stellar mass, $1.03~\Msun$. The constant mass evolution model (solid) burnt hydrogen with a radiative core whilst the other two (dashed and dotted) burnt hydrogen within a convective core and thus have larger core masses at the TAMS. Top: the evolution of the central degeneracy as a function of the core mass. Middle: the evolutionary time passed since the TAMS.
    Bottom: the rate of change of the core mass with respect to time. The tracks overlap in the core mass- central degeneracy parameter space due to the loss of thermal equilibrium on the Hertzsprung-Gap.}
    \label{fig:ImpactRadiativeHydrogenBurning}
\end{figure*}

Our method relies on the premise that all models are in thermal equilibrium, such that the mass history of a stellar model does not impact it's current properties.
However, it is known that stars evolve across the HG out of thermal equilibrium.
Once hydrogen is exhausted, the core no longer produces any luminosity and thus to be close to thermal equilibrium it must be approximately isothermal, otherwise energy diffuses outwards.
The maximum mass of an isothermal core is limited by the Sch{\"o}nberg-Chandrasekhar limit, $q_\mathrm{max} = M_\mathrm{c,max}/M \approx 0.1$ \citep{beechSchoenbergChandrasekharLimitPolytropic1988}.
Once $q_\mathrm{max}$ is exceeded, thermal equilibrium is lost and core contraction occurs on the thermal timescale in a quasi-static way, always maintaining a state very close to hydrostatic equilibrium.
However, if the core becomes electron degenerate, as it does in low mass stars ($M\lesssim 2~\Msun$), the Sch{\"o}nberg-Chandrasekhar limit no longer applies because degeneracy pressure can support the pressure exerted by the envelope.
Thus, degenerate cores can regain thermal equilibrium.

The impact of this can be seen by comparing the evolution of a star with a core that underwent radiative core-hydrogen burning with models at the same stellar mass ($M=1.03~\Msun$) but with cores that underwent convective core-hydrogen burning (Fig.~\ref{fig:ImpactRadiativeHydrogenBurning}).
The $\Mi = 1.03~\Msun$ model starts with a negligible core mass from radiative core-hydrogen burning and takes $\approx 2~\mathrm{Gyr}$ to reach the TAMS core masses of the convective core-hydrogen burning models with $\Mi = 1.38$ and $1.58~\Msun$.
WHen $\Mc\gtrsim 0.1~\Msun$, all stars are out of thermal equilibrium with non-isothermal cores.
In Fig.~\ref{fig:ImpactRadiativeHydrogenBurning} \textit{a}) we compare the evolution of the central degeneracy, $\etac$, as a function of the core mass, $\Mc$.
In thermal equilibrium, these tracks should not cross because the stellar structure should be uniquely defined by the combination of the stellar mass, core mass and central degeneracy.
Instead, we see that the efficiency of the hydrogen-burning shell is different despite models having the same $M$, $\Mc$ and $\etac$ as shown by the variety in the rate of change of the core mass in panel \textit{c}).
Due to the lack of thermal equilibrium, the lower TAMS core mass of the radiative core-hydrogen burning core leads to a difference in the shell-burning properties.
Once the cores become sufficiently degenerate, $\etac\gtrsim17$, thermal equilibrium is regained and the tracks converge to ignite helium at the same core mass and degeneracy.

The \textit{MINT} method is built on the assumption that the interpolation parameters uniquely define the stellar state.
Thus, at fixed stellar mass, the tracks in the time-proxy vs core mass parameter space should not cross because otherwise the derivatives of the interpolation parameters are not a unique function of the interpolation parameters.
In all models that undergo convective core-hydrogen burning, we find that the GB tracks do not cross and thus interpolation in this space is legitimate.
However, at each stellar mass we include a track which undergoes radiative core-hydrogen burning and thus starts with $\McTAMS\approx 0$.
This highlights the complexity of determining the properties on the HG without computing the full evolution from the TAMS.
However, we did not find this to cause any noticeable errors from our testing in section \ref{sec:MINTTesting}.

\subsection{Abundances of accreted material}

In this work we accrete material with the same composition as the surface composition.
Thus, the material is unprocessed and is not enhanced by processes in later stages of evolution such as dredge-up episodes.
However, companion stars may be enhanced in helium.
In addition, accretor stars can be polluted by helium-enhanced and CNO-processed material from the companion \citep{blaauwMassiveRunawayStars1993, 
izzardPopulationSynthesisBinary2009,
renzoEvolutionAccretorStars2021, el-badryUnicornsGiraffesBinary2022}.
Helium abundance has a significant impact on the envelope structure by altering both the mean molecular weight and opacity \citep[][and reference therein]{tannerHELIUMABUNDANCEOTHERCOMPOSITION2013}.
Increased helium envelope abundances could be accounted for by computing our model grid at a range of helium mass fractions as well as metallicities.
At low temperatures, CNO elements increase the metallicity and thus opacity of the envelope causing increased radii and lower effective temperatures \citep{marigoLowtemperatureGasOpacity2009,marigoLowTemperatureGasOpacities2022}, although this probably only impacts the AGB due to the low temperatures reached during this phase \citep{reeveComparativeStudyLowtemperature2023}.
Companion stars can also be enhanced by s-process elements such as barium \citep{aokiCarbonenhancedMetalpoorStars2007}, however these are in small enough abundance that they are unlikely to significantly impact the envelope structure.

\subsection{Computation time}

\begin{figure}
	% To  a figure from a file named example.*
 \centering
\includegraphics[width=\linewidth,height=0.9\textheight,keepaspectratio]{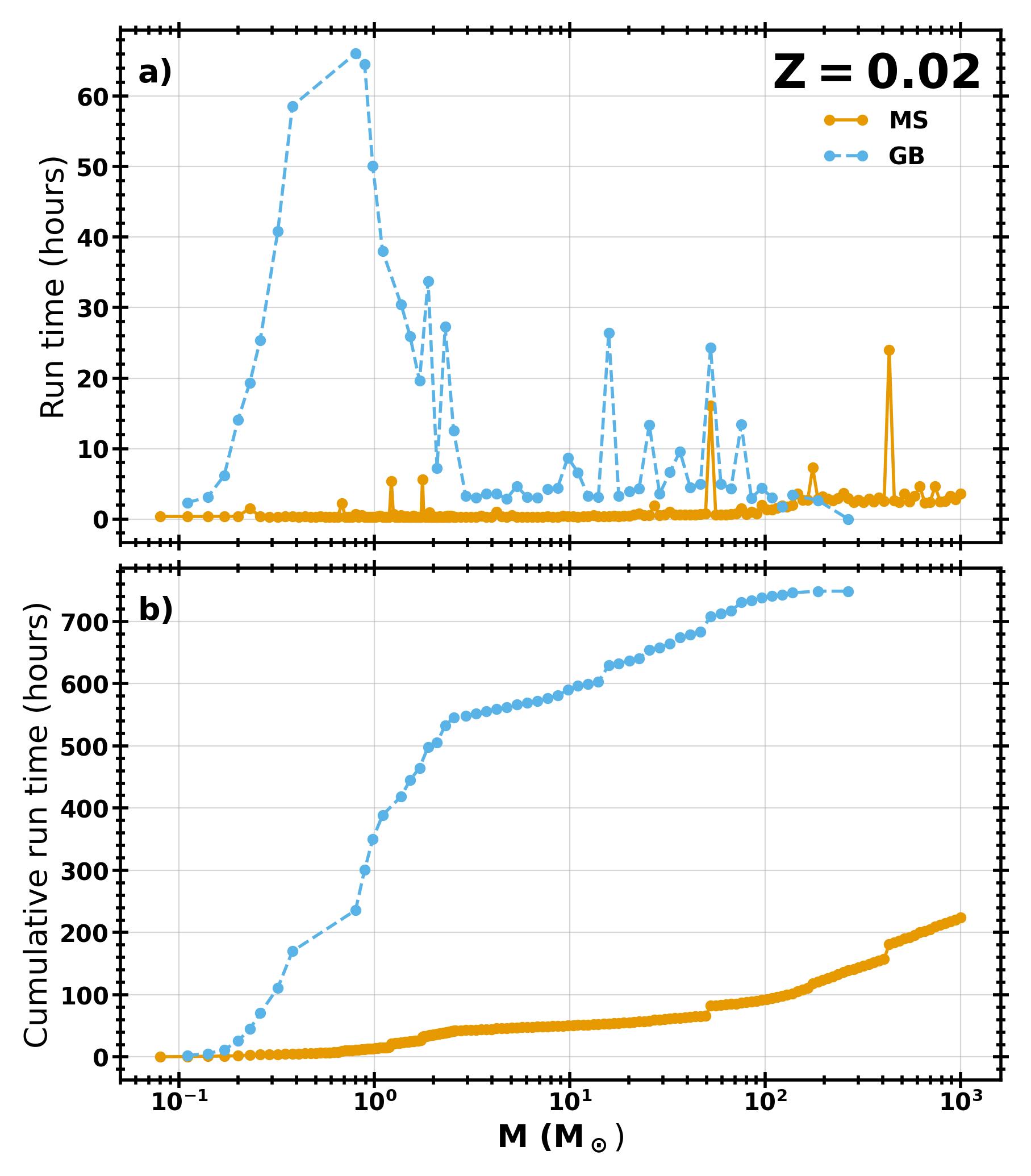}
    \caption{The \textit{MINT} grid run time for the MS (orange solid) and GB (blue dashed). \textit{a}) the run time as a function of stellar mass. \textit{b}) the cumulative run time for a total of 184 stellar masses on the MS and 54 stellar masses on the GB. The greatest contribution to run time is due to low-mass GB stars with degenerate helium ignition. Spikes are due to inefficiencies with the cluster used for this work and thus should be ignored. }
    \label{fig:RunTime}
\end{figure}

The running of \textit{MESA} grids requires the availability of many CPUs, thus some planning is required before trying to compute a number of grids at different metallicities.
We compute the run time for each evolutionary phase and total stellar mass to estimate the total amount of CPU time needed.
Note that \textit{MESA} is inherently multi-processed and we set the number of threads, \texttt{OMP\_NUM\_THREADS = 6}.
Increasing the number of threads for each \textit{MESA} run reduces the run time, however this is a non-linear relationship due to the cost associated with multi-threading.
Thus, when running a large number of \textit{MESA} models it is more CPU efficient to run more models at the same time, with a small number of threads used by each.
In Fig.~\ref{fig:RunTime} we compare the run time in hours for each evolutionary phase as a function of stellar mass.
Note that at each stellar mass, there are a number of \textit{MESA} models run with different initial core masses, with the exception of the MS.
The most computationally expensive phase is the low-mass GB due to degenerate helium ignition.
Note that spikes in the computation time are due to issues with particular nodes on the cluster used to compute our models, which cause the run time to be increased.
The total run time will also depend on the mass resolution used. 
Our calculation includes 184 stellar masses on the MS and 54 stellar masses on the GB over the range $0.08<M~/\Msun<1000$.

\section{Conclusions}
\label{sec:Conclusions}

We have produced a grid of models that cover the HG and RGB evolutionary phases for a range of core mass fractions. 
We thus include models with under-massive and over-massive cores that represent the resulting structures of mass transfer in binary systems, as well as strong stellar winds.
The model grid is used to produce an interpolation table for use in population synthesis codes. 
Population synthesis codes have traditionally been built using analytical formulae derived from single star evolution models.
This leads to problems when trying to model populations including binaries and higher order systems, because the underlying models do not contain the required information about stellar structures that have undergone significant mass changes.
This, and future, work is motivated by the aim to solve this problem, such that population synthesis codes can self consistently evolve multiple star systems, with the impacts of mass changes taken into account.
Stellar properties can be estimated from our table by interpolating as a function of the stellar mass, core mass fraction and the central degeneracy time-proxy.
We test our method by computing integrated tracks at constant mass in the range $M=1-16~\Msun$. 
We show that it successfully reproduces the HG and GB lifetime, core mass at helium ignition and the HR diagram.
The \textit{MINT} method is still under active development, including testing and expansion of the method to additional evolutionary phases and metallicities.

\section*{acknowledgments}

NRR thanks the EPSRC (EP/T518050) and Marion-Redfearn Trust for her funding grants. 
RGI acknowledges funding by the STFC grants ST/L003910/1 and
ST/R000603/1.
DDH thanks the UKRI/UoS for the funding grant H120341A.
We all thank Dr Giovanni Mirouh for his hard work in pioneering the early development of the \textit{MINT} project.

\newpage

\newpage

\bibliographystyle{mnras}
\bibliography{ZoteroLibrary}

\appendix

\section{MINT table data }
\label{append:MINTTable}

In all our models we save a number of useful quantities which make up the output columns of the interpolation table.
In Tables.~\ref{table:MINTColumnsScalar}-\ref{table:MINTColumnsVector} we list the columns included in \textit{MINT} with their corresponding descriptions.

\input{MINT_scalar_columns}

\input{MINT_vector_columns}

\end{document}

%% file: MINT_scalar_columns.tex
\begin{table*}
\centering
\caption{Scalar quantities included in the \textit{MINT} table for the evolutionary phase from core-hydrogen exhaustion to helium ignition.}
\label{table:MINTColumnsScalar}
\begin{tabular}{||l l||} 
\hline
 Column Name &
 Description \\
[0.5ex] 
\hline\hline
MASS & Total stellar mass ($\Msun$)\\
HELIUM\_CORE\_MASS\_FRACTION & Fractional He core mass ($\Mc/M$) \\
CENTRAL\_DEGENERACY & The electron chemical potential at the central mesh point, in units of $k_\mathrm{B}T$ (J)\\
RADIUS & Photospheric radius ($\Rsun$) \\ 
LUMINOSITY & Photospheric luminosity ($\Lsun$) \\
LUMINOSITY\_DIV\_EDDINGTON\_LUMINOSITY & Luminosity divided by the eddington luminosity \\
NEUTRINO\_LUMINOSITY & Power emitted in neutrinos, nuclear and thermal ($\Lsun$) \\
HELIUM\_LUMINOSITY & Total thermal power from the triple-alpha reaction, excluding neutrinos ($\Lsun$) \\
AGE & Model age (yr) \\
CENTRAL\_HYDROGEN & Central hydrogen mass fraction ($\Xc$)\\ 
CENTRAL\_HELIUM & Central helium mass fraction ($\Yc$) \\ 
HELIUM\_CORE\_RADIUS\_FRACTION & Relative core radius ($R_\mathrm{c}/R$) \\
CONVECTIVE\_CORE\_MASS\_FRACTION & Relative convective core mass ($M_\mathrm{c,conv}/M$) \\
CONVECTIVE\_CORE\_RADIUS\_FRACTION & Relative convective core radius ($R_\mathrm{c,conv}/R$) \\
CONVECTIVE\_CORE\_MASS\_OVERSHOOT\_FRACTION & Relative mass of main convective region in core including overshooting\\
CONVECTIVE\_CORE\_RADIUS\_OVERSHOOT\_FRACTION & Relative radius of main convective region in core including overshooting\\
CONVECTIVE\_ENVELOPE\_MASS\_FRACTION & Relative convective envelope mass ($M_\mathrm{env,conv}/M$) \\
CONVECTIVE\_ENVELOPE\_RADIUS\_FRACTION & Relative convective envelope radius (bottom envelope coordinate) \\
CONVECTIVE\_ENVELOPE\_MASS\_TOP\_FRACTION & Relative mass of the top of convective envelope \\
CONVECTIVE\_ENVELOPE\_RADIUS\_TOP\_FRACTION & Relative radius of the top of convective envelope \\
K2 & Apsidal constant \\
TIDAL\_E2 & Tidal E2 from Zahn \\
TIDAL\_E\_FOR\_LAMBDA & Tidal E for Zahn's lambda \\
MOMENT\_OF\_INERTIA\_FACTOR & $\beta^2$ from Claret AA 541, A113 (2012)= I/(MR${}^2$). \\
HELIUM\_CORE\_MOMENT\_OF\_INERTIA\_FACTOR & $\beta^2$ from Claret AA 541, A113 (2012)= I/(MR${}^2$) up to the helium core boundary \\
TIMESCALE\_KELVIN\_HELMHOLTZ & Stellar Kelvin-Helmholtz timescale (yr) \\
TIMESCALE\_DYNAMICAL & Stellar dynamical timescale (yr) \\
TIMESCALE\_NUCLEAR & Stellar nuclear timescale (yr) \\
MEAN\_MOLECULAR\_WEIGHT\_CORE & Mean molecular weight at central mesh point \\
MEAN\_MOLECULAR\_WEIGHT\_AVERAGE & Mean molecular weight average through star \\
FIRST\_DERIVATIVE\_CENTRAL\_DEGENERACY & First derivative of the central degeneracy with respect to time ($\mathrm{J}~\mathrm{yr}^{-1}$) \\
SECOND\_DERIVATIVE\_CENTRAL\_DEGENERACY & Second derivative of the central degeneracy with respect to time ($\mathrm{J}~\mathrm{yr}^{-2}$) \\
FIRST\_DERIVATIVE\_HELIUM\_CORE\_MASS\_FRACTION & First derivative of the helium core mass fraction with respect to time (yr${}^{-1}$) \\
SECOND\_DERIVATIVE\_HELIUM\_CORE\_MASS\_FRACTION & Second derivative of the helium core mass fraction with respect to time (yr${}^{-2}$). \\
WARNING\_FLAG & Warning flag = 1 if data warning, flag = 0 if data reliable \\
HELIUM\_IGNITED\_FLAG & Flag for the ignition of helium, = 1 if core helium burning started \\
INITIAL\_MASS & Initial mass used to form model, allows reconstruction of tracks \\
\hline
\end{tabular}
\end{table*}

%% file: MINT_vector_columns.tex
\begin{table*}
\centering
\caption{Vector quantities included in the \textit{MINT} table for the evolutionary phase from core-hydrogen exhaustion to helium ignition.}
\label{table:MINTColumnsVector}
\begin{tabular}{||l l||} 
\hline
 Column Name &
 Description \\
[0.5ex] 
\hline\hline
CHEBYSHEV\_MASS & Mass on Chebyshev grid ($\Msun$)\\
CHEBYSHEV\_TEMPERATURE & Temperature on Chebyshev grid (K) \\
CHEBYSHEV\_DENSITY & Density on Chebyshev grid ($\mathrm{g}~\mathrm{cm}^{-3}$)\\
CHEBYSHEV\_TOTAL\_PRESSURE & Total pressure on Chebyshev grid ($\mathrm{dyn}~\mathrm{cm}^{-2}$)\\
CHEBYSHEV\_GAS\_PRESSURE & Gas pressure on Chebyshev grid ($\mathrm{dyn}~\mathrm{cm}^{-2}$)\\
CHEBYSHEV\_RADIUS & Radius on Chebyshev grid ($\Rsun$)\\
CHEBYSHEV\_GAMMA1 & Adiabatic $\gamma_1$ on Chebyshev grid.\\
CHEBYSHEV\_PRESSURE\_SCALE\_HEIGHT & Pressure scale height on Chebyshev grid ($\Rsun$)\\
CHEBYSHEV\_DIFFUSION\_COEFFICIENT & Eulerian diffusion coefficient on Chebyshev grid ($\mathrm{cm}^2~\mathrm{s}^{-1}$)\\
CHEBYSHEV\_HELIUM\_MASS\_FRACTION & Helium-4 mass fraction on Chebyshev grid\\
CHEBYSHEV\_HYDROGEN\_MASS\_FRACTION & Hydrogen-1 mass fraction on Chebyshev grid \\

\hline
\end{tabular}
\end{table*}